\documentclass[preprint,aps,draft]{revtex4}

\usepackage{bm}% bold math

\begin{document}

\preprint{IZTECH-P-07-06}

\title{A way to get rid of cosmological constant and zero point energy 
problems 
of quantum fields through metric reversal symmetry}

\author{Recai Erdem}
\email{recaierdem@iyte.edu.tr}
\affiliation{Department of Physics,
{\.{I}}zmir Institute of Technology \\ 
G{\"{u}}lbah{\c{c}}e K{\"{o}}y{\"{u}}, Urla, {\.{I}}zmir 35430, 
Turkey} 

\date{\today}

\begin{abstract}
In this paper a framework is introduced to remove the  huge 
discrepancy between the empirical value of the cosmological constant and 
the contribution to the cosmological constant predicted from 
the vacuum energy of quantum fields. An extra dimensional space with 
metric reversal symmetry and $R^2$ gravity (that reduces to the usual R 
gravity after integration over extra dimensions) is considered to this 
end. The 
resulting 4-dimensional energy-momentum tensor (obtained after integration 
over extra dimensions) consists of terms that contain 
off-diagonally coupled pair of Kaluza-Klein modes. This, in turn, 
generically results in vanishing of the vacuum expectation value of the 
energy-momentum tensor for quantum fields, and offers a way to solve the 
problem of huge contribution of quantum fields to the vacuum energy 
density.
 \end{abstract}

\maketitle
\section{introduction}
The observation of the accelerated expansion of the universe 
\cite{Obs} boosted the studies on an old cosmological problem, namely, 
cosmological constant problem \cite{Weinberg}. The standard explanation 
for the accelerated expansion of the universe is a positive definite  
cosmological constant in Einstein field equations \cite{Carroll1, 
Carroll2}. A cosmological 
constant (CC) may be considered either as a geometrical object ( e.g. as 
the part of the curvature scalar that depends only on extra dimensions in 
a higher dimensional space) or as the energy density of a perfect fluid with 
negative pressure or a combination of both. 
(Although these two 
attributions may seem to be 
really two different manifestations of the same thing this distinction 
enables a more definite discussion of the problem as we shall see.) 
The vacuum expectation values of the energy-momentum tensors of quantum 
fields (i.e. the energy-momentum tensor due to zero modes of quantum 
fields) induce  energy-momentum tensors that has the form of the 
CC term in Einstein field equations. This 
identification is the main origin of the two (probably related ) most 
important cosmological 
constant problems; 1- why is the energy density ( $\sim (10^{-3}$ 
eV $)^4$ \cite{PDG} derived from the 
measurements of acceleration of the universe is so small compared to the 
energy scales associated with quantum phenomena ( that is, why is 
CC so small? ), 2- why does the zero modes of quantum 
fields contribute to the accelerated expansion of the universe so less 
than the expected?. 

There are many attempts, at least partially
to answer these questions, namely; symmetry principles, anthropic 
considerations, adjustment mechanisms, quantum cosmology and string 
landscape etc. \cite{Weinberg,Nobbenhuis}. None of these attempts have 
been wholly satisfactory. One of the main ideas proposed towards the 
solution of the problem is the use of symmetries such as supersymmetry and 
supergravity. However these symmetries are badly broken in nature. So 
it seems that they do not offer a viable solution. Recently a symmetry 
principle that does not suffer from such a phenomenological restriction 
was introduced \cite{Erdem1,Erdem2,Erdem3}. This symmetry amounts to 
invariance 
under the reversal of the sign of the metric and it has two different 
realizations. The first realization is implemented through the 
requirement of the invariance of physics under the multiplication of 
the coordinates by the imaginary number $i$ 
\cite{Erdem1,tHooft,Kamenshchik}. The second realization corresponds to 
invariance under signature reversal \cite{Bonelli,Erdem2,Duff} 
and may be realized through extra dimensional reflections\cite{Erdem2}.  
In this paper both realizations of the symmetry are named by a common 
name, "metric reversal symmetry". 
In the previous studies the symmetry is implemented for a cosmological 
constant that is geometrical in origin e.g. a bulk CC 
or a CC that is induced by the part of the curvature 
scalar that depends on the extra dimensions only. The aim of the present 
paper is to extend this symmetry to a possible contribution to CC 
induced by the vacuum expectation value of the energy-momentum tensor of 
quantum fields (i.e. quantum zero modes). The main difficulty in applying 
the symmetry to the contribution of the quantum zero modes 
is that, in the simple setting considered in the previous 
studies, it is not possible to impose it so that the matter Lagrangian 
corresponding to a field is non-vanishing after integration over extra 
dimensions (i.e. so that the field is observable at the usual 4-dimensions 
at the current accessible energies) while the quantum vacuum 
contributions of the fields are forbidden. This point will be mentioned 
in more detail in the following section. To this end, in this paper the 
space is taken to be a union of two $2(2n+1)$ dimensional spaces 
and the 
gravitational Lagrangian is taken to be $R^2$ where $R$ is the curvature 
scalar. Robertson-Walker metric is embedded in one of these $2(2n+1)$ 
dimensional spaces. Both realizations of the metric reversal symmetry
are imposed. The 4-dimensional 
Robertson-Walker metric reduces to the Minkowski metric 
after the symmetry imposed and 
the action corresponding to matter 
Lagrangian is forbidden by the requirement of the invariance under 
$x^A\,\rightarrow\,ix^A$. 
The requirement of the implementation of 
(either realization of) the 
symmetry on 
each space separately restricts the form of the gravitational action  
and only some part of the gravitational action 
survives and it can be identified by the usual Einstein-Hilbert action 
after integration over extra dimensions. After breaking the 
$x^A\,\rightarrow\,ix^A$ symmetry (while preserving the signature reversal 
symmetry) 
the Minkowski metric converts to the Robertson-Walker metric 
(with a slowly varying Hubble constant),
and results in a small non-vanishing matter Lagrangian (and action). The 
unbroken signature reversal symmetry imposes the
resulting matter Lagrangian generically contain at least one pair 
of off-diagonally coupled Kaluza-Klein modes in each homogeneous term 
and hence necessarily contains mixture of different 
Kaluza-Klein modes. This, in turn, causes the vacuum expectation value of 
energy-momentum tensor be zero as we shall see. Then the accelerated 
expansion of the universe may be attributed to some alternative methods 
such as quintessence \cite{quintessence,Copeland}, phantoms 
\cite{phantom,Copeland} etc. 
or a small 
CC may be induced classically after breaking of the 
$x^A\,\rightarrow\,ix^A$ symmetry as we shall see. 

\section{A Brief Overview of Metric Reversal Symmetry}

We consider two different realizations of a symmetry 
that reverses the sign of the metric
\begin{eqnarray}
ds^2\;=\;g_{AB}dx^A\,dx^B 
~\rightarrow 
~~-\,ds^2
\label{aa1} 
\end{eqnarray}
 and leaves the gravitational action
\begin{equation}
S_R = \frac{1}{16\pi\,G}\int \sqrt{(-1)^S g} \,R \,d^Dx \label{aa2}
\end{equation}
invariant, where $S$ and $g$ denote the number of space-like dimensions 
and determinant of the metric tensor, respectively. I call this symmetry, 
metric reversal symmetry.

The first realization of the symmetry \cite{Erdem1}is generated by the 
transformations
that multiply all coordinates by 
the imaginary number $i$
\begin{equation}
x^A\,\rightarrow\,i\,x^A~~,~~~g_{AB}\,\rightarrow\,g_{AB}~~. \label{aa3}
\end{equation}
The second realization \cite{Erdem2} is generated by the signature 
reversal
\begin{equation}
x^A\,\rightarrow\,x^A~~,~~~g_{AB}\,\rightarrow\,-g_{AB}~~. \label{aa4}
\end{equation}

The requirement of the invariance of Eq.(\ref{aa1}) under either of the 
realizations, Eq.(\ref{aa3}) and Eq.(\ref{aa4}) sets the dimension of the 
space $D$ to
\begin{equation}
D=2(2n+1)~~~,~~~~n=0,1,2,3,....~~~.
\label{aa5}
\end{equation}
Hence both realizations forbid a bulk cosmological constant (CC) term
\begin{equation}
S_C = \frac{1}{8\pi\,G}\int \sqrt{g} \,\Lambda \,d^Dx \label{aa6}
\end{equation}
(provided that $S_G$ remains invariant) where $\Lambda$ is the bulk 
CC. 

In fact these conclusions are valid for signature reversal symmetry in a 
more general setting where the 
whole space consists of a
$2(2n+1)$ dimensional subspace whose metric transforms like (\ref{aa4}) 
and the metric tensor for the rest of the space is even under the 
symmetry. In other words in a $D$-dimensional space where 
\begin{eqnarray}
&&x^A\,\rightarrow\,x^A~~,~~~g_{AB}\,\rightarrow\,-g_{AB}~~;~~~
A,B=0,1,2,3,5,....2(2n+1) ~~,
\label{aa7} \\
&&x^A\,\rightarrow\,x^A~~,~~~g_{A^\prime 
B^\prime}\,\rightarrow\,g_{A^\prime B^\prime}~~;~~~
A^\prime,B^\prime=2(2n+1)+1,2(2n+1)+2,.......,D 
\label{aa8} 
\end{eqnarray}
as well $S_G$ is allowed while $S_\Lambda$ is forbidden.

A higher dimensional metric with 
local Poincar\'{e} invariance may be written as 
\cite{Rubakov}
\begin{eqnarray}
ds^2\,=\,
\Omega(y^c)[
g_{\mu\nu}(x)
\,dx^{\mu}dx^\nu\,+\,
\tilde{g}_{\tilde{a}\tilde{b}}(y)\,dy^{\tilde{a}}dy^{\tilde{b}}]\,+\,
g_{e^\prime d^\prime}(y)\,dy^{e^\prime}dy^{d^\prime} 
\label{ab1} 
\end{eqnarray}
where $x$ and $\mu\,\nu\,=\,0,1,2,3$ denote the usual 
4-dimensional coordinates and indices; $y$ denotes extra dimensional 
coordinates, and 
$\tilde{a},\tilde{b}$=$4,5,...2(2n+1)$, 
$e^\prime,d^\prime$=$2(2n+1),....,D$ denote 
the extra dimensional indices. 
We let, 
\begin{eqnarray}
&&\Omega\,\rightarrow\,-\Omega 
~~,~~~g_{\mu\nu}\,\rightarrow\,g_{\mu\nu}
~~,~~~g_{\tilde{a}\tilde{b}}\,\rightarrow\,g_{\tilde{a}\tilde{b}} 
~~,~~~g_{e^\prime d^\prime}\,\rightarrow\,g_{e^\prime d^\prime}~~ . 
\label{ab3}
\end{eqnarray}
We take the underlying symmetry that induces (\ref{ab3}) be an extra 
dimensional reflection 
symmetry. For example one may take 
\begin{equation}
\Omega(y^c)\,=\,\cos k\,y~~~~~~~~y=y^D \label{ab4}
\end{equation}
where $k$ is some constant and take the symmetry transformation be a 
reflection about $kz=\frac{\pi}{2}$ given by
\begin{equation}
ky\,\rightarrow\,\pi\,-\,ky \label{ab5}~~ .
\end{equation}
There is a small yet important difference between 
simply postulating a signature reversal symmetry or realizing it through 
(\ref{ab1}) and (\ref{ab4}) 
although both forbid a cosmological constant (CC). 
In the case of 
(\ref{ab1}) and (\ref{ab4}), one may take a non-vanishing 
CC from the beginning and it cancels out after 
integration over extra dimensions while this is not possible if one 
simply postulates the metric reversal symmetry. 

The action functional corresponding to the matter sector is
\begin{equation}
S_M = 
\int \sqrt{(-1)^Sg} \,{\cal L}_M \,d^Dx \label{ac1}
\end{equation}
where ${\cal L}_M$ is the Lagrangian for a matter field. If the symmetry 
is applicable to the matter sector then the symmetry must leave $S_M$ 
invariant. One may take the 
dimension where the field propagates as $D=2(2n+1)$ so that (at least) the 
kinetic 
part of $S_M$ is invariant under the symmetry transformations. For 
example the kinetic part of the Lagrangian of a scalar field $\phi$ 
\begin{equation}
{\cal L}_{\phi\,k} 
= \frac{1}{2}g^{AB}\partial_A\phi\partial_B\phi
\label{ac2}
\end{equation}
transforms like 
$R$ under the transformations, (\ref{aa3}) and/or (\ref{aa4}) so that 
$S_M$ is invariant under the symmetry if $\phi$ propagates in a $2(2n+1)$ 
dimensional space and $\phi\rightarrow \pm\phi$ under the 
symmetry transformation. Meanwhile this allows non-zero contributions 
to the CC through the vacuum expectation of 
energy-momentum tensor of quantum fields. The 4-dimensional 
energy-momentum tensor for (\ref{ac2}) at low energies, 
$T_\mu^\nu$, 
is
\begin{eqnarray}
T_\mu^\nu\,=\int d^{D-4}y\,\Omega^{2n}
\sqrt{\tilde{g}\,g_e}\,
\{g^{\nu\tau}\partial_\tau\phi\partial_\mu\phi-
\frac{1}{2}\delta_\mu^\nu\,[
g^{\rho\tau}\partial_\rho\phi\partial_\tau\phi
\,+\,
\tilde{g}^{ab}\partial_a\phi\partial_b\phi
\,+\,
\Omega\,g^{ed}\partial_e\phi\partial_d\phi
]\} 
\label{ac3}
\end{eqnarray}  
where we employed the metric (\ref{ab1}), and $\tilde{g}$ and $g_e$ denote 
the determinants of $(\tilde{g}_{\tilde{a}\tilde{b}})$ and 
$(g_{e^\prime d^\prime})$, and $\delta_\mu^\nu$ 
denotes the Kronecker delta. 
If the signature reversal symmetry is imposed through an extra dimensional 
reflection, for example, by (\ref{ab4}) and (\ref{ab5}) then the last term 
in (\ref{ac3}) cancels out  while the other terms survive after 
the integration over the extra dimensions. So the 
4-dimensional energy-momentum tensor 
in general gives non-zero contribution to vacuum energy density through 
its vacuum expectation value after quantization.
One may allow
${\cal L}_{\phi\,k}$ 
by letting $\phi$ 
propagates in a $4n$ dimensional 
but this would allow a bulk CC. 
In other words one may adjust the dimension 
of the space where 
the field propagates so that (\ref{ac1}) is allowed and hence the symmetry 
is true for matter sector but this allows either a bulk CC or the 
contribution of quantum zero modes. 
The situation is 
the same for gauge fields and fermions. 
So one should 
consider this as a classical symmetry \cite{tHooft} or one should 
construct a more 
sophisticated framework where the symmetry applies both at classical 
and quantum levels. Constructing such a model will be the aim of the 
following sections.

\section{The need for both realizations of the symmetry and its 
implications}

The requirement of the isotropy and the homogeneity of the usual 
4-dimensional universe results in the metric
\begin{eqnarray}
ds^2&=&
\Omega(y)\,(\,dx_0^2\,-\,a(t)\,d\sigma^2\,)\,
+\,g_{ab}(y)\,dy^{a}dy^b 
\label{ba2} \\
&&y\,\equiv\, x_5=y_1,x_6=y_2,......,x_D=y_{D-4}
~~~~~~~a,b\,=1,2,3,.......,D-4 \nonumber \\
&&d\sigma^2\,=\,\frac{dr^2}{1-K^2r^2}+r^2d{\it \Omega}^2 ~~.
\nonumber
\end{eqnarray}
Further I impose the symmetry
\begin{eqnarray}
&&ds^2
~\rightarrow 
~~-\,ds^2~~~~~\mbox{as}~~~~~~
x^A\,\rightarrow\,i\,x^A~~,~~~g_{AB}\,\rightarrow\,g_{AB} \label{ba3} \\
&&~~~~~~A=0,1,2,3,5,....,D ~~~.\nonumber
\end{eqnarray}
This requires
\begin{eqnarray}
&&\Omega(y)\,\rightarrow\,\Omega(y) 
~~,~~~a(t)\,\rightarrow\,a(t)
~~,~~~
K^2r^2\,\rightarrow\,
K^2r^2~~,~~~
g_{ab}\,\rightarrow\,g_{ab}~~. \label{ba4}
\end{eqnarray}
This together with the requirement that after integration over extra 
dimensions it should correspond to the 
solution of the 4-dimensional Einstein equations with a cosmological 
constant (as the only 
source) implies that
\begin{equation}
a(t)\,=\,\mbox{constant}~~~,~~~~K^2=0 ~~. \label{ba5}
\end{equation}
In other words the first realization of the symmetry, Eq.(\ref{ba3}) 
requires the 4-dimensional part of the metric be the usual Minkowski 
metric, that is,
\begin{eqnarray}
ds^2&=&
\Omega(y)\,(
\,dx_0^2
\,-\,dx_1^2
\,-\,dx_2^2
\,-\,dx_3^2)
+\,g_{ab}(y)\,dy^{a}dy^b ~~.
\label{ba6} 
\end{eqnarray}
Eq.(\ref{ba6}) suggests that one may get rid of the problem of 
cosmological 
constant in the 4-dimensional cosmological constant (CC) (provided that 
extra dimensional contributions vanish) once the first realization of the 
metric reversal symmetry or (global) Poincar\'{e} symmetry is imposed. 
Then the smallness of the observational value of CC could be 
attributed to the breaking of the symmetry by a tiny amount if the 
renormalized value of CC due to vacuum fluctuations were in the order of 
the observed value of CC. On the other hand the renormalized value of CC 
is proportional to 
the particle masses \cite{renorm-cc}. So even a free electron contributes 
to CC by an amount that is $\sim 10^{33}$ times 
larger than the observational value of CC. Therefore the first realization 
of metric reversal 
symmetry by itself can not be used to make CC vanish 
(or tiny).  In the next section we will see how the signature reversal 
symmetry (realized through extra dimensional reflections) can be used to 
make the contribution of the quantum zero modes vanish. 
However the first 
realization has an advantage over the second one especially when the 
second realization is considered to be an extra dimensional reflection 
of the form of (\ref{ab5}). Extra dimensional reflections do not act on 
the 4-dimensional 
coordinates so they can not forbid a contribution from the 4-dimensional 
part of the metric, for example through $a(t)$ while the first realization 
always does by setting it to zero as we have seen. So in the next 
section we will employ both realizations of the symmetry. The second 
realization through extra dimensional reflections will cancel 
the contributions to CC while the first one will allow a small 
CC after it is broken by a small amount.

Next see what is the form of the conformal factor $\Omega$ when both 
realizations of the symmetry are imposed. We have obtained in (\ref{ba6}) 
the form of the metric after the first realization of the symmetry is 
imposed. Eqs.(\ref{ba3},\ref{ba4}) set the form of the 
conformal 
factor $\Omega$ 
in (\ref{ba2}) to 
one of the followings 
\begin{equation}
\Omega(y)\,=\,\Omega(|y|)~~~~~\mbox{or}~~~~~
\Omega(y)\,=\,f(y)f(iy)~~~~ (e.g. \cos{ky}\cosh{ky} )
\label{ba8}
\end{equation}
where $f(y)$ is an even function in $y$ i.e. $f(-y)=f(y)$. 
Next 
apply (\ref{ab5}) to (\ref{ba8}) and require 
(\ref{ab3}) and 
take the extra dimension $y$ be an $S^1/Z_2$ interval. 
This restricts the form of $\Omega$ to
\begin{equation}
\Omega(y)\,=\,\cos{k|y|}~~~~~\mbox{or}~~~~~
\Omega(y)\,=\,\tan{k|y|}
\label{ba9}
\end{equation}
where $\cot{k|z|}$ has been excluded because it blows out at the location 
of the branes at $k|y|=0$ and $k|y|=\pi$. For simplicity I take
\begin{equation}
\Omega(y)\,=\,\cos{k|y|}
\label{ba10}
\end{equation}
in the next section whenever necessary.

\section{The Model: Classical Aspects} 

In this section we employ both realizations of the metric reversal 
symmetry in a space that is the sum of two $2(2+1)$ 
dimensional spaces (where the usual 4-dimensional is embedded in one of 
them) and 
modify the curvature term $S_G$ so that the metric reversal symmetry 
becomes a good candidate to explain the huge discrepancy between the 
observed value of cosmological constant (CC) and the theoretically 
expected contribution to it through quantum zero modes. 
In this study I 
adopt the view that the symmetry forbids both the geometrical and the 
vacuum energy density contributions to CC. Hence CC is 
forced to be zero when the symmetry is manifest, and it is tiny when the 
symmetry is broken by a tiny amount (instead of seeking a solution where 
both contributions 
cancel each other up to a very big precession to explain the observed 
value of CC). In this 
section the main classical aspects of a framework to this end are
introduced.

Consider the whole space be a sum of two $2(2n+1)$ dimensional spaces with 
the metric
\begin{eqnarray}
ds^2&=&
g_{AB}dx^A\,dx^B 
\,+\,g_{A^\prime B^\prime}dx^{A^\prime}\,dx^{B^\prime} \nonumber \\
&=&
\Omega_z(z)[
g_{\mu\nu}(x)
\,dx^{\mu}dx^\nu\,+\,
\tilde{g}_{ab}(y)\,dy^{a}dy^b]
\,+\,
\Omega_y(y)\tilde{g}_{A^\prime B^\prime}(z)\,dz^{A^\prime}dz^{B^\prime} 
\label{c1} \\
&&\Omega_y(y)\,=\,\cos{k|y|})
~~,~~~\Omega_z(z)\,=\,\cos{k^\prime|z|} \label{c1a} \\
A,B&=&0,1,2,3,5,....N~~,~~~N=2(2n+1)~~,~~~~ 
A^\prime,B^\prime=1^\prime,2^\prime,,....N^\prime~~,~~~N^\prime=2(2m+1) 
\nonumber \\
&&~~~~~~~~~~~\mu\nu=0,1,2,3~,~~~a,b=1,2,...,N-4~~,~~~n,m=0,1,2,3......~~. 
\nonumber
\end{eqnarray}
The usual four dimensional space is embedded in the first space 
$g_{AB}dx^A\,dx^B$ 
as it is evident from (\ref{c1}). We take the action be invariant under 
both realizations of metric reversal symmetry, that is,
\begin{eqnarray}
&&ds^2
\,\rightarrow 
\,-\,ds^2~~~\mbox{as}~~~~
x^A\,\rightarrow\,i\,x^A\,,~~
x^{A^\prime}\,\rightarrow\,i\,x^{A^\prime}\,,~~
g_{AB}\,\rightarrow\,g_{AB}\,,~~ 
g_{A^\prime B^\prime}\,\rightarrow\,g_{A^\prime B^\prime} 
\label{c2} \\
&&\Rightarrow~~\Omega_z\,\rightarrow\,\Omega_z 
\,,~~\Omega_y\,\rightarrow\,\Omega_y 
\,,~~~g_{\mu\nu}\,\rightarrow\,g_{\mu\nu}
\,,~~\tilde{g}_{ab}\,\rightarrow\,\tilde{g}_{ab} 
\,,~~\tilde{g}_{A^\prime B^\prime}\,\rightarrow\,\tilde{g}_{A^\prime 
B^\prime} \label{c3}
\end{eqnarray}
and
\begin{eqnarray}
&&ds^2
\,\rightarrow 
\,-\,ds^2~~~\mbox{as}~~~~
ky\,\rightarrow\,\pi\,-\,ky~,~~ 
k^\prime z\,\rightarrow\,\pi\,-\,k^\prime z~,~~
x^A\,\rightarrow\,x^A\,,~~
x^{A^\prime}\,\rightarrow\,x^{A^\prime}
\label{c4} \\
&&\Rightarrow~~\Omega_z\,\rightarrow\,-\Omega_z 
\,,~~\Omega_y\,\rightarrow\,-\Omega_y 
\,,~~g_{\mu\nu}\,\rightarrow\,g_{\mu\nu}
\,,~~\tilde{g}_{ab}\,\rightarrow\,\tilde{g}_{ab} 
\,,~~\tilde{g}_{A^\prime B^\prime}\,\rightarrow\,\tilde{g}_{A^\prime 
B^\prime} ~~.\label{c5}
\end{eqnarray}
As in (\ref{ba6}) and (\ref{ba10}) the requirements of the homogeneity and 
isotropy of the 
4-dimensional space 
together with the equations (\ref{c2}-\ref{c5}) set $g_{\mu\nu}$ to the 
Minkowski metric $\eta_{\mu\nu}=diag(1,-1,-1,-1)$ and the conformal 
factors to (\ref{c1a}).

\subsection{Curvature Sector}
We replace the gravitational action in (\ref{aa2}) by an $R^2$ action
\begin{eqnarray}
S_R &=& \frac{1}{16\pi\,\tilde{G}}\int 
\,dV\,\tilde{R}^2 
\label{ca1} \\
&&dV\,=\,dV_1\,dV_2~,~~dV_1\,=\,\sqrt{g(-1)^S} \,d^Nx 
~,~~dV_2\,=\,
\sqrt{g^\prime
(-1)^{S^\prime}} 
\,d^{N^\prime}x^\prime
\label{ca2} \\
&&\tilde{R}=R(x,x^\prime)+R^\prime(x,x^\prime) \label{ca3}
\end{eqnarray}
where the 
unprimed quantities denote those corresponding to the 
$N=2(2n+1)$ 
dimensional space, and the
primed quantities denote those corresponding to the 
$N^\prime=2(2m+1)$ 
dimensional space. Under the transformations (\ref{c4},\ref{c5})
\begin{eqnarray}
&&dV_1\,\rightarrow \,-\,dV_1~,~~dV_2\,\rightarrow \,\,dV_2~~
~~~\mbox{as}~~~~
ky\,\rightarrow\,\pi\,-\,ky~,~~ 
x^A\,\rightarrow\,\,x^A\,,~~
x^{A^\prime}\,\rightarrow\,\,x^{A^\prime}
\label{ca4} \\
&&dV_1\,\rightarrow \,dV_1~,~~dV_2\,\rightarrow \,-\,dV_2~~
~~~\mbox{as}~~~~
k^\prime z\,\rightarrow\,\pi\,-\,k^\prime z~,~~
x^{A}\,\rightarrow\,\,x^{A}~,~~
x^{A^\prime}\,\rightarrow\,\,x^{A^\prime}
\label{ca5} \\
&&R\,\rightarrow \,R~,~~R^\prime\,\rightarrow \,-R^\prime
~~~\mbox{as}~~~~
ky\,\rightarrow\,\pi\,-\,ky~,~~ 
x^A\,\rightarrow\,\,x^A\,,~~
x^{A^\prime}\,\rightarrow\,\,x^{A^\prime}
\label{ca6} \\
&&R\,\rightarrow \,-R~,~~R^\prime\,\rightarrow \,R^\prime
~~~\mbox{as}~~~~
k^\prime z\,\rightarrow\,\pi\,-\,k^\prime z~,~~
x^A\,\rightarrow\,\,x^A\,,~~
x^{A^\prime}\,\rightarrow\,\,x^{A^\prime}~~.
\label{ca7}
\end{eqnarray}
We observe that 
\begin{eqnarray}
&&dV\,=\,dV_1dV_2\,\rightarrow \,-\,dV \label{ca8} \\
&&R^2\,
\rightarrow \,
R^2
~,~~
R^{\prime 2}\,
\rightarrow \,
R^{\prime 2}~,~~
R\,R^\prime\,
\rightarrow \,
-R\,R^\prime\,
\label{ca9}
\end{eqnarray}
under the action of the symmetry transformations to only one of the 
spaces, the unprimed or the primed spaces. So, only the cross terms 
$RR^\prime$ are allowed. In other words only these terms may survive after 
integration over extra dimensions. In fact it 
is obvious from the above transformation rules that an Einstein-Hilbert 
type of action is not allowed directly because each piece $R$ and 
$R^\prime$ in $\tilde{R}$ is odd while $dV$ is even under a 
transformation applied to both
subspaces, the unprimed and the primed subspaces.
Since only $RR^\prime$ terms are allowed (\ref{ca1}) 
becomes 
\begin{eqnarray} S_R &=& \frac{M^{N+N^\prime-4}}{16\pi\,\tilde{G}}\int 
\sqrt{(-1)^S g} 
\sqrt{(-1)^{S^\prime} g^\prime} 
\,2\,R(x)\,R^\prime(x^\prime) 
\,d^Nx
\,d^{N^\prime}x^\prime \nonumber \\
&=& 
 \frac{1}{16\pi\,G}
\int 
\sqrt{(-1)^S g} 
\,R(x)\, \,d^Nx
\label{ca10} 
\end{eqnarray}
where
\begin{equation}
 \frac{1}{16\pi\,G}\,=\, 
 M_{pl}^2(\frac{M}{M_{pl}})^2M^{N+N^\prime-6}
\frac{1}{16\pi\,\tilde{G}}\int 
\sqrt{(-1)^{S^\prime} g^\prime} 
\,2\,R^\prime(x^\prime) 
\,d^Dx^\prime \label{ca11}
\end{equation}
and $\tilde{G}$ is a dimensionless constant. In other words in the usual 
4-dimensions at low energies (\ref{ca1}) 
is the same as the Einstein-Hilbert action (\ref{aa2}). The Newton's 
constant in $N$ dimensions, $G$ is related to the Newton's constant in 
$N+N^\prime$ 
dimensions  through Eq.(\ref{ca11}). The integral in (\ref{ca11}) is at 
the order of 
$\sim L^{N^\prime-2}\sim \frac{1}{M^{N^\prime -2}}$. Hence Eq.(\ref{ca11}) 
may explain the smallness of 
gravitational interaction compared to the other interaction if the 
energy scale of 
$L^\prime$ 
is much smaller than the Planck mass $M_{Pl}$ 
i.e. if 
$L^\prime>>\frac{1}{M_{Pl}}$ as in the models with large extra dimensions 
especially when $L(L^\prime)\,<\,\frac{1}{M}$.

\subsection{Matter Sector}

In this subsection we consider the matter action
\begin{eqnarray}
S_M &=& 
\int 
\,dV\,{\cal L}_M 
\label{cb1} \\
&&dV\,=\,\sqrt{(-1)^S g}\sqrt{(-1)^{S^\prime} g^\prime} \,d^Dx\,d^Dx^\prime
\nonumber
\end{eqnarray}
and we consider the 4-dimensional form of $S_M$ after integration over 
extra dimensional spaces. Then we study the vacuum expectation value of 
the energy-momentum tensor induced by the corresponding Lagrangian 
in the section after the next section.

It is evident that
under the first realization of the symmetry
\begin{eqnarray}
&&dV
\,\rightarrow \,\,
dV
~~~\mbox{as}~~~~
x^{A(A^\prime)}
\,\rightarrow\,
i\,x^{A(A^\prime)}~,~~
g_{AB(A^\prime B^\prime})\,\rightarrow\,g_{AB(A^\prime B^\prime)} 
\label{cb2} 
\end{eqnarray}
for a space consisting of the sum of two $2(2n+1)$ dimensional spaces as 
in (\ref{c1}). The kinetic part of ${\cal L}_M$ is not invariant under the 
transformations 
$x^{A(A^\prime)}\,\rightarrow\,i\,x^{A(A^\prime)}$ 
for the usual fields 
\cite{tHooft}. So $S_M$ is not invariant under the symmetry generated by 
$x^{A(A^\prime)}\,\rightarrow\,i\,x^{A(A^\prime)}$. In other words the 
first realization of the metric reversal symmetry is maximally broken in 
the matter sector (and hence the scale factor $a(t)$ in the 
Robertson-Walker metric may be time dependent). On the other hand I take a 
higher 
dimensional version of 
the $PT$ symmetry $x^{A(A^\prime)}\,\rightarrow\,-\,x^{A(A^\prime)}$ be 
almost exact and broken by a tiny amount. In other words I adopt
\begin{equation}
x^A\,\rightarrow\,-\,x^A ~~,~~~~
x^{A^\prime}\,\rightarrow\,-\,x^{A^\prime} 
\label{cb3}
\end{equation} 
which is a 
subgroup of the group generated by 
\begin{eqnarray} 
x^{A(A^\prime)}\,&\rightarrow&\,i\,x^{A(A^\prime)}\,
\rightarrow\,i\,(i\,x^{A(A^\prime)})\,
=-\,x^{A(A^\prime)}
\rightarrow\,i\,(i\,(i\,x^{A(A^\prime)}))
=-i\,x^{A(A^\prime)} \nonumber \\
&&\rightarrow\,i(i\,(i\,(i\,x^{A(A^\prime)})))
\,=\,x^{A(A^\prime)}~~.
\label{cb4}
\end{eqnarray}
The symmetries in (\ref{cb3}) are imposed on each subspace separately.
Next I impose an additional 4-dimensional PT symmetry generated by
\begin{equation}
x\,\rightarrow\,-x \label{cb5}~~.
\end{equation}
Eqs.(\ref{cb4},\ref{cb5}) 
together imply that a PT symmetry in the 4-dimensions 
and an additional PT-like symmetry in the extra dimensional sector are 
assumed. One observes that ${\cal L}_{SM}$ is invariant under 
Eqs.(\ref{cb4},\ref{cb5})  because 
$S_M$ and $dV$ are invariant under these symmetries. The eigenvectors of 
Eqs.(\ref{cb4},\ref{cb5}) do not mix because the Lagrangian ( so the 
Hamiltonian) is invariant under these symmetries. So the fields $\phi$ in 
the Lagrangian should be eigenvectors of these symmetries.

To make the argument more concrete
consider the Fourier decomposition (i.e. Kaluza-Klein decomposition) 
of a general field $\phi$ (where possible spinor or vector indices are 
suppressed).  For simplicity we take 
$\tilde{g}_{ab}=-\delta_{ab}$,
$g_{A^\prime B^\prime}=-\delta_{A^\prime B^\prime}$, and consider only the 
Fourier decomposition of $\phi$ corresponding to single dimensions $y$ and 
$z$ from each of the subspaces, the unprimed and the primed ones. We show 
that the Fourier expansions given below are the eigenvectors of
Eqs.(\ref{cb4},\ref{cb5}),
\begin{eqnarray}
\phi_{AA}(x,y,z)&=&\sum_{n,m} \,\phi^{AA}_{n,m}(x)\,
\sin{(n\,ky)}
\,\sin{(m\,k^\prime z)}
\label{cb6} \\
\phi_{AS}(x,y,z)&=&\sum_{n,m} \,\phi^{AS}_{n,m}(x)\,
\sin{(n\,ky)}
\,\cos{(m\,k^\prime z)}
\label{cb7} \\
\phi_{SA}(x,y,z)&=&\sum_{n,m} \,\phi^{SA}_{n,m}(x)\,
\cos{(n\,ky)}
\,\sin{(m\,k^\prime z)}
\label{cb8} \\
\phi_{SS}(x,y,z)&=&\sum_{n,m} \,\phi^{SS}_{n,m}(x)\,
\cos{(n\,ky)}
\,\cos{(m\,k^\prime z)}
\label{cb9} \\
&&
k=\frac{\pi}{L}~,~
k^\prime=\frac{\pi}{L^\prime}~,~~
0\leq\,y\,\leq\,L 
~,~~0\leq\,z\,\leq\,L^\prime~~,~~~n,m=0,1,2,..... \nonumber
\end{eqnarray}
where we have used 
$k=\frac{\pi}{L}$,
$k^\prime=\frac{\pi}{L^\prime}$ since
$0\leq\,y\,\leq\,L$,
$0\leq\,z\,\leq\,L^\prime$. In the case of fermions the integers 
$n$, $m$ in (\ref{cb6},\ref{cb9}) should be 
replaced by $\frac{1}{2}n$, $\frac{1}{2}m$, respectively.
One observes that
\begin{equation}
n(m)\,\rightarrow\,-n(m)~~~\mbox{as}~~~y(z)\,\rightarrow\,-y(z)
\label{cb10}
\end{equation}
since $n(m)$ are the eigenvalues of 
$\frac{\partial}{\partial y}\,
(\frac{\partial}{\partial z})$ i.e. they are the momenta corresponding to 
the 
directions $y$ and $z$. There are two eigenvalues i.e. $\pm\,1$ of the 
each transformation in (\ref{cb10}) since application of the 
transformations twice results in the identity transformation.

Now we show that the fields (\ref{cb6},\ref{cb9}) are the eigenstates of 
the transformations (\ref{cb10}). First consider (\ref{cb6}). Applying the 
transformation 
(\ref{cb3}) and using (\ref{cb10}), $\phi_{AA}$ in (\ref{cb6}) transforms 
to
\begin{eqnarray}
\phi_{AA}(x,y,z)
&\rightarrow&\phi^\prime(x,y^\prime,z)
\,=\,\sum_{n,m} \,\phi^{AA}_{-n,m}(x)\,
\sin{(n\,ky)}
\,\sin{(m\,k^\prime z)}~~~\mbox{as}~~~~y\,\rightarrow -y
\label{cb11} \\
&\rightarrow&\phi^\prime(x,y^,z^\prime)
\,=\,\sum_{n,m} \,\phi^{AA}_{n,-m}(x)\,
\sin{(n\,ky)}
\,\sin{(m\,k^\prime z)}~~~\mbox{as}~~~~z\,\rightarrow -z~~.
\label{cb12} 
\end{eqnarray}
There will be no mixture of the 
eigenstates of (\ref{cb3}) in the Lagrangian
because the Lagrangian 
is invariant under (\ref{cb3}). So $\phi_{AA}$ is either odd or 
even under (\ref{cb3}). 
In the light of 
(\ref{cb10},\ref{cb12}) 
the eigenstates of $\phi_{AA}$ 
under the 
transformation are determined by 
$\phi^{AA}_{n,m}(x)$. The same conclusion is true for all 
$\phi$'s (\ref{cb6},\ref{cb9}). 
So, for all $\phi$'s (\ref{cb6},\ref{cb9}) we have 
two cases for each symmetry in (\ref{cb10})
\begin{eqnarray}
&&\phi_{-n,m}(-x)\,=\,
\pm\phi_{-n,m}(x)\,=\,
\pm\phi_{n,m}(x) \label{cb13} \\
&&\phi_{n,-m}(-x)\,=\,
\pm\phi_{n,m}(x)\,=\,
\pm\phi_{n,m}(x) ~~.\label{cb14} 
\end{eqnarray}
Meanwhile one may write 
(\ref{cb6},\ref{cb9}) in the following form as well
\begin{eqnarray}
\phi_{AA}(x,y,z)&=&\frac{1}{2}\sum_{n,m} (
\,\phi^{AA}_{n,m}(x)\,
-\,\phi^{AA}_{-n,m}(x)\,)
\sin{(n\,ky)}
\,\sin{(m\,k^\prime z)} \nonumber \\
&=&\frac{1}{2}\sum_{n,m} (
\,\phi^{AA}_{n,m}(x)\,
-\,\phi^{AA}_{n,-m}(x)\,)
\sin{(n\,ky)}
\,\sin{(m\,k^\prime z)}
\label{cb15}\\ 
\phi_{AS}(x,y,z)&=&\frac{1}{2}\sum_{n,m} (
\,\phi^{AS}_{n,m}(x)\,
-\,\phi^{AS}_{-n,m}(x)\,)
\sin{(n\,ky)}
\,\cos{(m\,k^\prime z)} \nonumber \\
&=&\frac{1}{2}\sum_{n,m} (
\,\phi^{AS}_{n,m}(x)\,
+\,\phi^{AS}_{n,-m}(x)\,)
\sin{(n\,ky)}
\,\cos{(m\,k^\prime z)}
\label{cb16}\\ 
\phi_{SA}(x,y,z)&=&\frac{1}{2}\sum_{n,m} (
\,\phi^{SA}_{n,m}(x)\,
+\,\phi^{SA}_{-n,m}(x)\,)
\sin{(n\,ky)}
\,\sin{(m\,k^\prime z)} \nonumber \\
&=&\frac{1}{2}\sum_{n,m} (
\,\phi^{SA}_{n,m}(x)\,
-\,\phi^{SA}_{n,-m}(x)\,)
\cos{(n\,ky)}
\,\sin{(m\,k^\prime z)}
\label{cb17}\\ 
\phi_{SS}(x,y,z)&=&\frac{1}{2}\sum_{n,m} (
\,\phi^{SS}_{n,m}(x)\,
+\,\phi^{SS}_{-n,m}(x)\,)
\cos{(n\,ky)}
\,\cos{(m\,k^\prime z)} \nonumber \\
&=&\frac{1}{2}\sum_{n,m} (
\,\phi^{SS}_{n,m}(x)\,
+\,\phi^{SS}_{n,-m}(x)\,)
\cos{(n\,ky)}
\,\cos{(m\,k^\prime z)}~~.
\label{cb18} 
\end{eqnarray}
It is evident from Eq.(\ref{cb15},\ref{cb18}) that 
$\phi_{AA}$ 
is antisymmetric under both of 
$n\rightarrow\,-n$,
$m\rightarrow\,-m$,
$\phi_{AS}$
is antisymmetric under 
$n\rightarrow\,-n$
while it is symmetric under
$m\rightarrow\,-m$, 
$\phi_{SA}$
is symmetric under 
$n\rightarrow\,-n$
while it is antisymmetric under
$m\rightarrow\,-m$, and 
$\phi_{SS}$ is symmetric under both
of $n\rightarrow\,-n$,
$m\rightarrow\,-m$. This result will be important in the value of $S_M$ 
after integration over extra dimensions.

\subsubsection{Scalar Field}

First consider ${\cal L}_{\phi k}$, 
the kinetic part of the Lagrangian 
${\cal L}_{Mk}$ for a scalar field (in 
the space given in (\ref{c1}))
\begin{eqnarray}
&&{\cal L}_{\phi\,k} \,=\,
{\cal L}_{\phi\,k1}\,+\, 
{\cal L}_{\phi\,k2} 
\label{cba1} \\
&&{\cal L}_{\phi\,k1}\,=\, 
\frac{1}{2}g^{AB}\partial_A\phi\partial_B\phi~~,~~~
{\cal L}_{\phi\,k2} 
\,=\,\frac{1}{2}g^{A^\prime B^\prime}
\partial_{A^\prime}\phi\partial_{B^\prime}\phi~~.
\label{cba2}
\end{eqnarray}

Once the breaking of the first realization of the symmetry 
in the matter sector  is 
granted we may go on to seek the implications of the 
manifestations of the residual 
symmetry (\ref{cb3},\ref{cb5}) and the 
second realization of the symmetry 
given by Eqs.(\ref{c4},\ref{c5}) that remains unbroken. 
${\cal L}_M$ (i.e. ${\cal L}_{\phi k}$ in this case) is 
even under 
the simultaneous application of the signature reversal symmetry to  
both subspaces because $dV$ is even 
under the symmetry and we require the invariance of $S_M$ (i.e. $S_{\phi 
k}$ in this case). 
So any $\phi$ may be 
written as a sum of the eigenstates of the symmetry. The eigenvalues of 
the symmetry transformation 
$k^{(\prime)}y(z)\,\rightarrow\,
\pi\,-\,k^{(\prime)}y(z)$ are $\pm\,1$ because application of the 
transformation twice results in the identity transformation. Because 
$g^{AB}(g^{A^\prime B^\prime})$ is odd then the terms 
$\partial\phi\partial\phi$ are odd as well under the symmetry 
transformation. So the kinetic term in (\ref{cba1}) contains mixed 
eigenstates of the symmetry. In the following paragraphs we will 
identify these eigenstates with odd and even terms in the Fourier 
decomposition (i.e. Kaluza-Klein decomposition) of $\phi$. Then this 
result 
will have important consequences in the following paragraphs.
In the next paragraph we see, through an example, 
explicitly how 
$S_M$ 
contains 
mixing of different Kaluza-Klein modes off-diagonally. This result, in 
turn, will be crucial in ensuring vanishing of the vacuum expectation 
value of energy-momentum tensors of quantum fields in the section 
after the next section.

To illustrate the idea I avoid unnecessary 
complications and consider the 
simplest realistic case; $N=6$, $N^\prime=2$. The kinetic part of $S_M$ 
(i.e. $S_{\phi}$ in this case) for $\phi_{SS}$ of Eq.(\ref{cb9})
in the space (\ref{c1}) where the conformal factors are of the form 
(\ref{c1a}) is given by ( see Appendix A)
\begin{eqnarray}
S_{\phi k} &=& 
\frac{1}{8}(LL^\prime)^2\int\,d^4x\,\{
4\partial_\mu 
[\,\phi_{1,2}(x)\,+\,\phi_{1,0}(x)]
\,\partial_\nu(\,\phi_{0,0}(x)\,)\nonumber \\
&&+\, 
4\partial_\mu 
[\,\phi_{0,2}(x)\,+\,\phi_{0,0}(x)
\,+\,\phi_{2,2}(x)\,+\,\phi_{2,0}(x)]
\,\partial_\nu(\,\phi_{1,0}(x)\,)\nonumber \\
&&+\, 
4\eta^{\mu\nu}\sum_{r=1,s=1}^{\infty}
\partial_\mu 
[\,\phi_{|r-1|,|s-2|}(x)\,+\,\phi_{|r-1|,s+2}(x)
\nonumber \\
&&+\,2\phi_{|r-1|,s}(x)\,+\,\phi_{r+1,|s-2|}(x)\,+\,\phi_{r+1,s+2}(x)
\,+\,2\phi_{r+1,s}(x) 
\,] \partial_\nu(\,\phi_{r,s}(x)\,) 
\nonumber \\
&&\,-4k^2\,\,
\sum_{r=1,s=0}\,r[\,(|r-1|)
(\,\phi_{|r-1|,|s-2|}(x)\,+\,\phi_{|r-1|,s+2}(x)
\,+\,2\,\phi_{|r-1|,s}(x)\,)
\nonumber\\
&&+\,(r+1)
(\,\phi_{r+1,|s-2|}(x)\,+\,\phi_{r+1,s+2}(x)
\,+\,2\,\phi_{r+1,s}(x)\,)\,-\,\phi_{r+1,s}(x)\,)\,]
\phi_{r,s}(x) \nonumber \\
&&\,-4\frac{1}{2}k^{\prime 2}\,\,
\sum_{r=0,s=1}\,s
\,[\,(|s-3|)\phi_{r,|s-3|}(x)\,+\,(s+3)\,\phi_{r,s+3}(x) \nonumber \\
&&+\,3(|s-1|)\,\phi_{r,|s-1|}(x)\,+\,3(s+1)(\,\phi_{r,s+1}(x)\,]
\,\phi_{r,s}(x) \}~~.
\label{cba6} 
\end{eqnarray}
The expressions for 
$\phi_{AS}$, $\phi_{SA}$, $\phi_{AA}$ 
are the same as (\ref{cba5}) up to 
minus and pluses in front of the $\phi_{mn}$ terms. Hence the expressions 
for $\phi_{AS}$, $\phi_{SA}$, $\phi_{AA}$ 
are the same as (\ref{cba6}) because the change in the sign of 
the 
coefficients of $\phi_{mn}$ are compensated by the change of the sign 
due to the symmetry properties of  $\phi_{mn}$'s  under $n\rightarrow -n$
$m\rightarrow -m$. Although the expressions for $S_{\phi k}$ for all
$\phi_{AA}$, $\phi_{AS}$, $\phi_{SA}$, $\phi_{SS}$ are essentially the 
same and given by (\ref{cba6}), in fact the $S_{\phi k}$ for $\phi_{SS}$ 
has an important difference than the others because only that 
result contains the zero mode $\phi_{0,0}$ that is identified by the usual 
particles. So I take 
$\phi_{SS}$ as the only physically relevant state for $\phi$. 
One observes that Eq.(\ref{cba6}) contains only off-diagonal mixing of 
Kaluza-Klein modes. One may 
easily see that a bulk mass term for $\phi$ results in essentially the 
same form as the 4-dimensional kinetic term in (\ref{cba6}) where the 
derivatives are absent. Any other power of $\phi$ necessarily 
contains off-diagonal mixings of Kaluza-Klein modes. These 
observations are important when the vacuum expectation of 
energy-momentum tensor is obtained to give zero in the exact 
manifestation of extra dimensional reflection symmetry. A 
more detailed analysis of Eq.(\ref{cba6}) and these points will be given 
in the next section.

Next consider a bulk mass term (for $\phi_{SS}$)
\begin{eqnarray}
S_{\phi m} &=& 
\frac{1}{2}m\int 
\,\sqrt{(-1)^S g}\sqrt{(-1)^{S^\prime} g^\prime} \,d^Dx\,d^Dx^\prime
\phi^2 \nonumber \\
&=& 
\frac{1}{2}m\,LL^\prime\int\,d^4x\,\{
\sum_{n,m,r,s}
\,\phi_{n,m}(x)\,
\phi_{r,s}(x)\nonumber \\
&&\int_0^L\,dy\,
\cos{k y}\, 
\cos{(n\,k|y|)}
\cos{(r\,k|y|)}
\int_0^{L^\prime}\,dz\,
\cos^3{k^\prime z}
\cos{(m\,k^\prime|z|)})
\,\cos{(s\,k^\prime|z|)}) \nonumber \\
&=&
\frac{1}{64}m(LL^\prime)^2\int\,d^4x\,\{
\sum_{n,m,r,s}\,\phi_{n,m}(x)\,\phi_{r,s}(x)
[(\delta_{n,-r-1}+\delta_{n,1-r})
+\delta_{n,r-1}+\delta_{n,1+r})
\nonumber \\
&&
\times\,(\delta_{m,-s-3}+
\delta_{m,3-s}+
\delta_{m,s-3}+
\delta_{m,s+3} \nonumber \\
&&+
3\delta_{m,-s-1}+
3\delta_{m,1-s}
+3\delta_{m,s-1}+
3\delta_{m,s+1})]\} ~~.\label{cba7}
\end{eqnarray}

The common aspect of the equations (\ref{cba6}) and {\ref{cba7}) are that 
the Kaluza-Klein modes mix in such a way that there are no diagonal terms 
i.e. the terms of the form 
$\phi_{n,m}\phi_{n,m}$. 
In fact this is a 
generic property of all possible terms for all kinds of fields i.e. 
scalars, fermions, gauge fields or any other kind of field. All terms 
necessarily contain at least a pair of Kaluza-Klein modes that couple in 
a non-diagonal way. This can be seen as follows: A pair of fields that 
mix in a diagonal way (i.e. as 
$\phi_{n,m}\phi_{n,m}$) is even under either of the transformations in 
(\ref{c4}) since 
it corresponds to the terms of the form $\cos^2{n\,ky}\sin^2{m\,k^\prime 
z}$. If the whole terms consists of such pairs then the whole term is even 
under (\ref{c4}). However the volume element is odd under 
either of the transformations in (\ref{c4}). 
So such a term can not exist i.e. it must contain at least one pair of 
fields that couple in a off-diagonal way. This fact plays a crucial role 
in making the vacuum expectation value of the energy momentum tensor zero 
in the exact manifestation of the metric reversal symmetry. In the next 
subsection we consider one additional example, that is, the kinetic term 
for fermions because it is not a straightforward generalization of 
the scalar 
case. We will see that the same conclusion also holds in that case 
as expected.

\subsubsection{Fermionic Fields}

The kinetic term of the Lagrangian for fermionic fields in the space given 
by (\ref{c1}) in the presence of the signature 
reversal symmetry (where the conformal factors and the unprimed space are 
given by (\ref{c1a}) and (\ref{ba6}) ) is
\begin{eqnarray}
{\cal L}_{fk}\,=\,
i\bar{\psi}\Gamma^{A}\partial_A\psi 
\,+\,i\bar{\psi}\Gamma^{A^\prime}\partial_{A^\prime}\psi~~.
\label{cbc1}
\end{eqnarray} 
For simplicity I take
\begin{equation} 
g_{\mu\nu}=\eta_{\mu\nu}~,~~ 
\tilde{g}_{ab}=-\delta_{ab}~,~~ 
\tilde{g}_{A^\prime B^\prime}=-\delta_{A^\prime B^\prime} ~~.\label{cbc2}
\end{equation}
In fact $g_{\mu\nu}=\eta_{\mu\nu}$ is enforced by the symmetry, the 
4-dimensional homogeneity and isotropy of the metric as we have discussed 
in the previous section. So 
\begin{eqnarray}
\Gamma^A&=&(\cos{\frac{k\,z}{2}}\tau_3
\,+\,i\sin{\frac{k\,z}{2}}\tau_1)^{-1} \otimes\gamma^A 
\nonumber \\
\Gamma^{A^\prime}&=&(\cos{\frac{k\,y}{2}}\tau_3
\,+\,i\sin{\frac{k\,y}{2}}\tau_1)^{-1}\otimes\gamma^{A^\prime} 
\label{cbc3}
\end{eqnarray}
where
\begin{eqnarray}
\{\Gamma^{A(A^\prime)},\Gamma^{B(B^\prime)}\}=2g^{AB(A^\prime 
B^\prime)}~~~,~~~~
\{\gamma^A,\gamma^B\}=2\eta^{AB}~~~,~~~~
\{\gamma^{A^\prime},\gamma^{B^\prime}\}=-2\delta^{A^\prime,B^\prime}
\label{cbc4}
\end{eqnarray}
and $\tau_3$, $\tau_1$ are the diagonal and the off diagonal real Pauli 
matrices, and $\otimes$ denotes tensor product. In the case of fermions 
one should use the complex expansion for the Fourier expansion 
\begin{eqnarray}
\psi(x,y,z)&=&\sum_{n,m} \,\psi_{n,m}(x)\,
e^{\frac{i}{2}\,n\,ky}
\,e^{\frac{i}{2}\,m\,k^\prime z}\nonumber \\
&=&\sum_{n,m} \,(\,
\psi^{nS}_{n,m}(x)\,\cos{(\frac{1}{2}n\,ky)}
\,+\,
\psi^{nA}_{n,m}(x)\,\sin{(\frac{1}{2}n\,ky)}\,)
\,e^{\frac{i}{2}\,m\,k^\prime z}\nonumber \\
&=&\sum_{n,m} \,(\,
(\,\psi^{mS}_{n,m}(x)\,\cos{(\frac{1}{2}m\,k^\prime z)}
\,+\,
\psi^{mA}_{n,m}(x)\,\sin{(\frac{1}{2}m\,k^\prime z)})
e^{\frac{i}{2}\,n\,ky}
\label{cbc5} 
\end{eqnarray}
where
\begin{eqnarray}
&&\psi^{nS}_{n,m}(x)\,=\,
\frac{1}{2}(\,\psi_{n,m}(x)\,+\,\psi_{-n,m}(x)\,)~,~~
\psi^{nA}_{n,m}(x)\,=\,
\frac{i}{2}(\,\psi_{n,m}(x)\,-\,\psi_{-n,m}(x)\,) \nonumber \\
&&\psi^{mS}_{n,m}(x)\,=\,
\frac{1}{2}(\,\psi_{n,m}(x)\,+\,\psi_{n,-m}(x)\,)~,~~
\psi^{nA}_{n,m}(x)\,=\,
\frac{i}{2}(\,\psi_{n,m}(x)\,-\,\psi_{n,-m}(x)\,) ~~.\label{cbc6}
\end{eqnarray}
Next we substitute (\ref{cbc5}) in (\ref{cbc1}) to get $S_{fk}$.
To be specific we take 
$N=6$ and $N^\prime=2$ as in the previous subsubsection. Then (\ref{cbc1}) 
becomes (see Appendix B)
\begin{eqnarray}
S_{fk}
&=& 
\frac{1}{32}(LL^\prime)^2\int\,d^4x\,\{
\sum_{n,m,r,s} 
\,[\,
i\psi_{n,m}(x)\,\tau_3\otimes\gamma^\mu\partial_\mu(\,\psi_{r,s}(x)
\nonumber\\
&&\times\, 
(\delta_{n,r+2}+\delta_{n,r-2})
(\delta_{m,s+5}+
\delta_{m,s-3}+
2\delta_{m,s+1}
(\delta_{m,s-5}+
\delta_{m,s+3}+
2\delta_{m,s-1}) \nonumber \\
&& -\psi_{n,m}(x)\,\tau_3\otimes\gamma^\mu\partial_\mu(\,\psi_{r,s}(x)
\nonumber\\
&&\times\, 
(\delta_{n,r+2}+\delta_{n,r-2})
(\delta_{m,s+3}+
\delta_{m,s-5}-
\delta_{m,s+5}-\delta_{m,s-3}+
2\delta_{m,s-1}
-2\delta_{m,s+11}) \nonumber \\
&&-\,
\frac{1}{2}\psi_{n,m}(x)\,(r-n)\,\tau_3\otimes\gamma^y \psi_{r,s}(x)
\nonumber\\
&&\times\, 
(\delta_{n,r+2}+\delta_{n,r-2})
(\delta_{m,s+3}+
\delta_{m,s-5}-
\delta_{m,s+5}-\delta_{m,s-3}+
2\delta_{m,s-1}
-2\delta_{m,s+11}) \nonumber \\
&&+\,\frac{1}{2}
(r-n)\,
\psi_{n,m}(x)\,\tau_1\otimes\gamma^y
\,\psi_{r,s}(x) \nonumber \\
&&\times\, 
(\delta_{n,r+2}+\delta_{n,r-2})
(\delta_{m,s+3}+
\delta_{m,s-5}-
\delta_{m,s+5}-\delta_{m,s-3}+
2\delta_{m,s-1}
-2\delta_{m,s+11}) \nonumber \\
&&-\,
\frac{1}{2}\psi_{n,m}(x)\,(s-m)\,\tau_3\otimes\gamma^y \psi_{r,s}(x) 
\nonumber \\
&&\times\, 
(\delta_{n,r+1}+\delta_{n,r-1})
(\delta_{m,s+6}+\delta_{m,s-6}
+3\delta_{m,s+2}+3\delta_{m,s-2})
\nonumber \\
&&+\,(s-m)\frac{1}{2}
\,
\psi_{n,m}(x)\,\tau_1\otimes\gamma^y
\,\psi_{r,s}(x) \nonumber \\
&&\times\, (\delta_{n,r-1}-\delta_{n,r+1})
(\delta_{m,s+6}+\delta_{m,s-6}+3\delta_{m,s+2}+3\delta_{m,s-2})\}
\label{cbc7}
\end{eqnarray} 
 where we have used the identity 
$\cos{u}\,(\cos{\frac{u}{2}}\tau_3\,+\,i\sin{\frac{u}{2}}\tau_1)^{-1}$=
$(\cos{\frac{u}{2}}\tau_3\,+\,i\sin{\frac{u}{2}}\tau_1)$.
We see that, in this case as well, each homogeneous term consists of 
one off-diagonally coupled pair of Kaluza-Klein modes.

\section{the relation to Linde's model}
It is 
evident from (\ref{cba6}) that the 4-dimensional kinetic term contains 
the zero mode $\phi_{00}$ while the other terms i.e the mass terms do 
not contain the zero mode. This implies that there is a zero mass 
eigenstate 
that contains $\phi_{00}$. However the form of (\ref{cba6}) is rather 
involved since it involves, in general, mixing of all Kaluza-Klein modes.
An important aspect of this mixing 
is the absence of diagonal terms in the mixing 
terms. We will see in the next section how this plays a crucial role in 
making the vacuum expectation value of energy-momentum tensor zero. Before 
passing to this issue, first we should make the form of (\ref{cba6}) more 
manageable. In any case one should diagonalize 
(\ref{cba6}) so that, at least, the 
fields in the 4-dimensional kinetic term couple to each other diagonally 
i.e. we should pass to the interaction basis. One observes due to the 
signature reversal symmetry (induced through extra dimensional 
reflections) that all the 
terms in the 4-dimensional kinetic term in (\ref{cba6}) are mixed so 
that 
the terms with odd $n$'s mix with the even $n$'s, 
and the odd $m$'s with odd $m$'s, 
the even $m$'s with even $m$'s. There is the same behavior for the terms 
with the coefficient $k^2$, and a similar behavior for the terms with the 
coefficient $k^{\prime 2}$ (the odd $n$'s mix with the odd $n$'s,
the even $n$'s mix with the even $n$'s, and the odd $m$'s mix with the 
even $m$'s and vice versa). So the form given by the 
4-dimensional part of (\ref{cba6}) may be only 
induced by the mixture of either of 
\begin{eqnarray}
&&\phi^{OO}_{SS}(x,y,z)\,=\,\sum_{j,l=0} 
\,\phi^{OOSS}_{2j+1,2l+1}(x)\,
\cos{(2j+1)\,ky}
\,\cos{(2l+1)\,k^\prime z} \nonumber \\
&&\mbox{and} \nonumber \\
&&\phi^{EO}_{SS}(x,y,z)\,=\,\sum_{j,l=0} 
\,\phi^{EOSS}_{2j,2l+1}(x)\,
\cos{(2j)\,ky}
\,\cos{(2l+1)\,k^\prime z}
\label{lba7} 
\end{eqnarray}
or
\begin{eqnarray}
&&\phi^{EE}_{SS}(x,y,z)\,=\,\sum_{j=1,l=0} 
\,\phi^{EESS}_{2j,2l}(x)\,
\cos{(2j)\,ky}
\,\cos{(2l)\,k^\prime z} \nonumber \\
&& \mbox{and} \nonumber \\
&&\phi^{OE}_{SS}(x,y,z)\,=\,\sum_{j,l=0} 
\,\phi^{OESS}_{2j+1,2l}(x)\,
\cos{(2j+1)\,ky}
\,\cos{(2l)\,k^\prime z} ~~.
\label{lba8} 
\end{eqnarray}
The each sum may be an infinite series if all modes are mixed 
or it may correspond to a set of finite sums if the modes mix with each 
other in a set of subsets of $r$ and $s$ in (\ref{cba6}). In the expansion 
of $\phi^{EE}_{SS}$ the sum over $j$ starts from one 
because we take the zero mode $\phi_{00}$ in a different eigenstate as 
we will see. The 
requirement that the internal symmetries that may be 
induced by extra dimensional symmetries and the usual space-time 
symmetries are independent requires the whole space be a direct product of 
the 4-dimensional space with the extra dimensional space. This, in turn, 
requires all $\phi_{n,m}(x)$'s in the above equations be the same up to 
constant coefficients, that is,
\begin{equation}
\phi^{XY}_{SS,n,m}\,=\,C_{n,m}^{XYSS}\phi^{XY}(x) \label{lba9}
\end{equation}
where $X,Y$ may take the values $O$, $E$, and $C_{n,m}^{XYSS}$ is some 
constant with the condition that it leads to a finite series. For example, 
one may take 
\begin{equation}
C_{n,m}=\frac{|n-2|\,|m-2|}{(n^2+1)(m^2+1)} \label{lba91}
\end{equation}
where 
$|n-2|\,|m-2|$ 
is included to make the analysis of the zero mass eigenstate more 
manageable as will see . Then Eqs.(\ref{lba7},\ref{lba8}) become 
\begin{eqnarray}
\phi^{OO}(x,y,z)&=&[\sum_{j,l=0} \,
C^{OO}_{2j+1,2l+1}\,\cos{(2j+1)ky}\,\cos{(2l+1)k^\prime z}]
\phi^{OO}(x) \nonumber \\
&=&
[\sum_{j,l=0} \,
\frac{
|2j-1|\,|2l-1|
}{((2j+1)^2+1)((2l+1)^2+1)}\,
\cos{(2j+1)ky}\,\cos{(2l+1)k^\prime z}]\phi^{OO}(x)\nonumber \\
&\mbox{and}& \nonumber \\
\phi^{EO}(x,y,z)&=&[\sum_{j,l=0} \,
C^{EO}_{2j,2l+1}\,\cos{(2j)ky}\,\cos{(2l+1)k^\prime z}]
\phi^{EO}(x)\nonumber \\
&=&[\sum_{j,l=0} \,
\frac{
|2j-2|\,|2l-1|
}{((2j)^2+1)((2l+1)^2+1)}\,
\cos{(2j)ky}\,\cos{(2l+1)k^\prime z}]\phi^{EO}(x)\,
\label{lba10} 
\end{eqnarray}
or
\begin{eqnarray}
\phi^{EE}(x,y,z)&=&\sum_{j=1,l=0} \,
C^{EE}_{2j,2l}\,\cos{(2j)ky}\,\cos{(2l)k^\prime z}]
\phi^{EE}(x)\nonumber \\
&=&[\sum_{j,l=0} \,
\frac{
|2j-2|\,|2l-2|
}{((2j)^2+1)((2l)^2+1}\,
\cos{(2j)ky}\,\cos{(2l)k^\prime z}]\phi^{EE}(x)\nonumber \\
&\mbox{and}& \nonumber \\
\phi^{OE}(x,y,z)&=&\sum_{j,l=0} \,
C^{OE}_{2j+1,2l}\,\cos{(2j+1)ky}\,\cos{(2l)k^\prime z}
\phi^{OE}(x)\nonumber \\
&=&[\sum_{j,l=0} \,
\frac{
|2j-1|\,|2l-2|
}{((2j+1)^2+1)((2l)^2}+1)\,
\cos{(2j+1)ky}\,\cos{(2l)k^\prime z}]\phi^{OE}(x)\,
\label{lba11} 
\end{eqnarray}
where the $SS$ indices are suppressed. In the light of 
(\ref{lba10},\ref{lba11}) Eq.(\ref{cba6}) becomes
\begin{eqnarray}
S_{\phi k} &=& 
\frac{1}{2}(LL^\prime)^2\int\,d^4x\,\{
2\eta^{\mu\nu}\partial_\mu (\,\phi_{1,0})
\partial_\nu(\,\phi_{0,0})
\,+\,2C_1\,C_2
\eta^{\mu\nu}\partial_\mu 
(\phi^{EO}(x)\,)\,
\partial_\nu
\,(\phi^{OO}(x)\,) \nonumber \\
&&+\,
2C_3\,C_4
\eta^{\mu\nu}\partial_\mu \,
(\phi^{EE}(x)\,)\,
\partial_\nu
\,(\phi^{OE}(x)\,) \nonumber \\
&&\,-k^2\,\,
[\,2C_5\,C_6
\,\phi^{OO}(x)\,
\phi^{EO}(x) \,+\,2C_7\,C_8
\,\phi^{EE}(x)\,
\phi^{OE}(x) \,]\nonumber \\
&&\,-\frac{1}{2}k^{\prime 2}\,
[\,2C_9\,C_{10}
\,\phi^{OO}(x)\,
(\phi^{OE}(x) \nonumber \\
&&
+\,2C_{11}\,C_{12}
\,\phi^{EE}(x)\,\phi^{EO}(x)\,] \,\}
\label{lba12} 
\end{eqnarray}
where the form of the coefficients $C_i$, $i=1,2,3,...,12$ are given in 
Appendix C.
The diagonalization of (\ref{lba12}) results in
\begin{eqnarray}
S_{\phi k} &=& 
\frac{1}{2}(LL^\prime)^2\int\,d^4x\,\{
\eta^{\mu\nu}(\partial_\mu\phi_1)
\partial_\nu(\,\phi_1)
\,-\,\eta^{\mu\nu}(\partial_\mu\phi_2)
\partial_\nu(\,\phi_2) \nonumber \\
&&+\,C_1\,C_2\,(\,
\eta^{\mu\nu}(\partial_\mu \phi_3(x)\,)\,
(\partial_\nu\phi_3(x)\,)
\,-\,\eta^{\mu\nu}
\,\partial_\mu(\phi_4(x)\,)\,
\partial_\nu(\phi_4(x)\,)\,)
\nonumber \\
&&+\,C_3\,C_4\,(\,
\eta^{\mu\nu}\partial_\mu(\phi_5(x)\,)\,\partial_\nu (\phi_5(x)\,) 
\,-\,
\eta^{\mu\nu}\partial_\mu (\phi_6(x)\,)\,
\partial_\nu\,(\phi_6(x)\,)\,)\,] 
\nonumber \\
&&-k^2\,
[\,C_5\,C_6\,(\,\phi_3(x)\,\phi_3(x)
\,-\,
\phi_4(x)\,\phi_4(x)\,) 
\nonumber \\
&&+\,C_7\,C_8\,(\,\phi_5(x)\,\phi_5(x)
\,-\,
\phi_{6}(x)\,\phi_{6}(x)\,) \,]\nonumber \\
&&-\frac{1}{2}k^{\prime 2}\,
[\,C_9\,C_{10}\,(\,\phi_{7}(x)\,\phi_{7}(x)
\,-\,
\phi_{8}(x)\,\phi_{8}(x)\,)
 \nonumber \\
&&+\,C_{11}\,C_{12}
\,(\,\phi_{9}(x)\,\phi_{9}(x)\,-\,\phi_{10}(x)\,\phi_{10}(x)\,)
\,] \,\}
\label{lba14} 
\end{eqnarray}
where
\begin{eqnarray}
&&\phi_1\,=\,\phi_{0,0}\,+\,\phi_{1,0}~~~,~~~~
\phi_2\,=\,\phi_{0,0}\,-\,\phi_{1,0}
~~~,~~~~
\phi_3\,=\,\phi^{EO}\,+\,\phi^{OO} 
~~~,~~~~
\phi_4\,=\,\phi^{EO}\,-\,\phi^{OO} 
\nonumber \\
&&\phi_5\,=\,\phi^{EE}\,+\,\phi^{OE} 
~~~,~~~~
\phi_6\,=\,\phi^{EE}\,-\,\phi^{OE} 
~~~,~~~~
\phi_7\,=\,\phi^{OO}\,+\,\phi^{OE} 
~~~,~~~~
\phi_8\,=\,\phi^{OO}\,-\,\phi^{OE} 
\nonumber \\
&&\phi_9\,=\,\phi^{EE}\,+\,\phi^{EO} 
~~~,~~~~
\phi_{10}\,=\,\phi^{EE}\,-\,\phi^{EO} ~~.\label{lba15}
\end{eqnarray}

It is evident from (\ref{lba14}) that the scalar kinetic Lagrangian 
(\ref{cba6}) is equivalent to a Lagrangian that consists of a set of 
usual scalars and a set of ghost scalars. In fact this conclusion is valid 
for all quadratic terms for all fields e.g. 
$\bar{\psi}_{n,m}\psi_{r,s}$ 
where $n\neq r$ and/or $m\neq s$ due to the symmetry and this term is 
equivalent to $\frac{1}{2}(\bar{\psi}_1\psi_1-\bar{\psi}_2\psi_2)$ where 
$\bar{\psi}_1= \psi_{n,m}+\psi_{r,s}$, 
$\bar{\psi}_2=\psi_{n,m}-\psi_{r,s}$. This setting is similar to 
Linde's model \cite{Linde} and its variants \cite{LindeVar}. Only mixing 
between the usual particles and ghost sector may be induced through 
quartic and higher order terms. A detailed analysis of such 
possible mixings and suppressing these couplings needs a separate study by 
its own.

\section{Vacuum expectation value of energy-momentum tensor in the 
presence of metric reversal symmetry}

The 4-dimensional energy momentum tensor corresponding to the 
action (\ref{lba12}) is
\begin{eqnarray}
T^\nu_\mu&=&\frac{2}{\sqrt{(-1)^Sg}
\sqrt{(-1)^{S^\prime}g^\prime}}g_{\mu\rho}\frac{\delta\,S_M}
{\delta\,g_{\nu\rho}}\,=\,
2\partial_\mu 
\,\phi_{1,0}(x)
\,\partial^\nu\,\phi_{0,0}(x) \nonumber \\
&&+\, 
\,+\,2C_1\,C_2
\partial_\mu 
(\phi^{EO}(x)\,)\,
\partial^\nu
\,(\phi^{OO}(x)\,) 
\,+\,
2C_3\,C_4
\partial_\mu [\,
(\phi^{EE}(x)\,)\,
\partial^\nu
\,(\phi^{OE}(x)\,) \nonumber \\
&&\,-\,\delta_\mu^\nu\,\{\,
\eta^{\mu\nu}\partial_\mu (\,\phi_{1,0})
\partial_\nu(\,\phi_{0,0})
\,+\,C_1\,C_2
\eta^{\mu\nu}\partial_\mu 
(\phi^{EO}(x)\,)\,
\partial_\nu
\,(\phi^{OO}(x)\,) \nonumber \\
&&+\,
C_3\,C_4
\eta^{\mu\nu}\partial_\mu \,
(\phi^{EE}(x)\,)\,
\partial_\nu
\,(\phi^{OE}(x)\,) \nonumber \\
&&-k^2\,\,
[\,C_5\,C_6
\,\phi^{OO}(x)\,
\phi^{EO}(x) \,+\,C_7\,C_8
\,\phi^{EE}(x)\,
\phi^{OE}(x) \,]\nonumber \\
&&\,-\frac{1}{2}k^{\prime 2}\,
[\,C_9\,C_{10}
\,\phi^{OO}(x)\,
(\phi^{OE}(x) 
\,+\,C_{11}\,C_{12}
\,\phi^{EE}(x)\,\phi^{EO}(x)\,] \,\}~~.
\label{d1}
\end{eqnarray} 
It is evident from (\ref{d1}) that all terms consist of  
off-diagonally coupled Kaluza-Klein modes. As we have remarked before any 
4-dimensonally Lagrangian term (after integration over extra dimensions) 
necessarily contains at least a pair of Kaluza-Klein modes that are 
off-diagonally coupled in the space given by (\ref{c1}). (As we have 
remarked in the previous section, this is due to the 
fact that if a term wholly consists of pairs of diagonally coupled 
Kaluza-Klein modes then that term is even under the signature reversal 
symmetry in contradiction with the invariance of the action under the 
signature reversal symmetry.) This, in turn, leads to cancellation of the 
vacuum expectation value of $T^\nu_\mu$ since it is proportional to terms 
of the form
\begin{equation}
<0|T^\nu_\mu|0>~\propto~~
<0|\,a_{n,m}a_{r,s}^\dagger|0>\,=\,0~,~~
<0|\,a_{r,s}^\dagger 
a_{r,s}|0>\,=\,0
~~~~~~~n\neq r~~~~\mbox{and/or}
~~~~m\neq s
\label{d2}
\end{equation}
(because $a_{r,s}|0>\,=\,0$, and $[\,a_{n,m}, 
a_{r,s}^\dagger\,]\,=\,0$ for 
$n\neq r$ and/or $m\neq s$) where 
$a_{n,m}$, $a_{n,m}^\dagger$ are the creation and annihilation operators 
in the expansion the quantum fields (in Minkowski space) given by
\begin{equation}
\phi_{n,m}(x)\,=\,\sum_{\vec{k}}\,[\,
a_{n,m}(\vec{k})\,e^{-iEt}
e^{i\vec{k}.\vec{x}}
\,+\,a_{n,m}^\dagger(\vec{k})\,e^{iEt}
e^{-i\vec{k}.\vec{x}}\,]~~.
\label{d3}
\end{equation}
The same reasoning is true for all fields. Therefore the vacuum energy 
density of all fields in this scheme is zero.

In this scheme the Casimir effect can be seen as follows: Introduction of 
(metallic) boundaries into the vacuum results in a change in the vacuum 
configuration for the usual particles while the ghost sector vacuum 
remains the same. This point can be seen better when one considers the 
the energy momentum tensor written in terms of 
the usual and ghost fields by using (\ref{lba14})
\begin{eqnarray}
T^\nu_\mu&=&
(\partial_\mu\phi_1(x)
\partial^\nu\phi_1(x))
\,-\,
\partial_\mu\phi_2(x)
\partial^\nu\phi_2(x)) \nonumber \\
&&+\,C_1\,C_2\,(\,
\partial_\mu \phi_3(x)
\partial^\nu\phi_3(x)
\,-\,
\partial_\mu\phi_4(x)
\partial^\nu\phi_4(x)\,)
\nonumber \\
&&+\,C_3\,C_4\,(\,
\partial_\mu\phi_5(x)\partial^\nu\phi_5(x)\,) 
\,-\,
\partial_\mu\phi_6(x)
\partial^\nu\phi_6(x)\,) 
\nonumber \\
&&-\,\frac{1}{2}\delta_\mu^\nu\,\{\,
\eta^{\mu\nu}(\partial_\mu\phi_1)
\partial_\nu(\,\phi_1)
\,-\,\eta^{\mu\nu}(\partial_\mu\phi_2)
\partial_\nu(\,\phi_2) \nonumber \\
&&+\,C_1\,C_2\,(\,
\eta^{\mu\nu}(\partial_\mu \phi_3(x)\,)\,
(\partial_\nu\phi_3(x)\,)
\,-\,\eta^{\mu\nu}
\,\partial_\mu(\phi_4(x)\,)\,
\partial_\nu(\phi_4(x)\,)\,)
\nonumber \\
&&+\,C_3\,C_4\,(\,
\eta^{\mu\nu}\partial_\mu(\phi_5(x)\,)\,\partial_\nu (\phi_5(x)\,) 
\,-\,
\eta^{\mu\nu}\partial_\mu (\phi_6(x)\,)\,
\partial_\nu\,(\phi_6(x)\,)\,)\,] 
\nonumber \\
&&-k^2\,
[\,C_5\,C_6\,(\,\phi_3(x)\,\phi_3(x)
\,-\,
\phi_4(x)\,\phi_4(x)\,) 
\,+\,C_7\,C_8\,(\,\phi_5(x)\,\phi_5(x)
\,-\,
\phi_{6}(x)\,\phi_{6}(x)\,) \,]\nonumber \\
&&-\frac{1}{2}k^{\prime 2}\,
[\,C_9\,C_{10}\,(\,\phi_{7}(x)\,\phi_{7}(x)
\,-\,
\phi_{8}(x)\,\phi_{8}(x)\,)
 \nonumber \\
&&+\,C_{11}\,C_{12}
\,(\,\phi_{9}(x)\,\phi_{9}(x)\,-\,\phi_{10}(x)\,\phi_{10}(x)\,)
\,] \,\}~~.
\label{d4}
\end{eqnarray}
To see the situation better let us consider a simple case, for example the 
part of the energy-momentum tensor that contains the zero mode. After 
introduction of the (metallic) boundary the vacuum expectation value of 
the 
corresponding part of the energy momentum tensor changes as follows
\begin{eqnarray}
<0|\,T^\nu_\mu\,|0>_0&=&
<0|\,(\partial_\mu\phi_1)
\partial^\nu(\,\phi_1)\,|0>_0
\,-\,
<0|\,(\partial_\mu\phi_2)
\partial^\nu(\,\phi_2)\,|0>_0\,=\,0\,\rightarrow\,
<0|\,T^\nu_\mu\,|0>_{\Sigma_1} \nonumber \\
&=&
<0|\,(\partial_\mu\phi_1)
\partial^\nu(\,\phi_1)\,|0>_{\Sigma_1}
\,-\,
<0|\,(\partial_\mu\phi_1)
\partial^\nu(\,\phi_1)\,|0>_0\,
\neq\,0
\label{d5}
\end{eqnarray}
where the subscript $0$ denotes complete vacuum (without any boundary) and 
the subscript $\Sigma_1$ denotes the vacuum in the presence of the 
(metallic) boundaries. It is evident that this scheme results in an 
automatic application of the usual subtraction prescription in 
the calculation of Casimir energies i.e an automatic subtraction of the 
zero point energy from the total vacuum energy in the presence of a 
boundary.

To summarize; I have shown that the quantum zero modes do not contribute 
to cosmological constant (CC) in the scheme presented here in the presence 
of metric reversal symmetry. Now, for the sake of completeness, I discuss 
the other possible contributions to CC. The first 
additional contribution is a bulk CC (that is geometric in 
origin). The transformations (\ref{ca4}) and/or (\ref{ca5}) (or 
equivalently the form of the conformal factors given in (\ref{c1a}) ) 
forbid a bulk CC (or equivalently make it vanish 
after integration over extra dimensions).  The second possible 
contribution is a 4-dimensional CC that may be induced by 
the part of the scalar curvature that depends only on extra 
dimensions. Eq.(\ref{ca10}) implies that such a 
contribution vanishes provided that the half of the extra dimensions in 
the $2(2n+1)$ dimensional space (embedding the usual 4-dimensional space) 
are spacelike and half are timelike as in \cite{Erdem1}. The next possible 
contribution is the vacuum energy induced by the vacuum expectation value 
of Higgs field, and  is about $\sim 10^{55}$ times the 
observational value of CC. This contribution has the 
form of a bulk CC, and hence vanishes provided that 
Higgs field propagates in the whole space or in its a $2(2k+1)$ 
dimensional subspace. Another possible standard contribution is the vacuum 
expectation value of the QCD vacuum (that is about $10^{44}$ times the 
observational value of CC). 
At classical level the same condition as Higgs 
field applies to the space where the corresponding condensate forms. 
However a rigorous conclusion needs an analysis at quantum 
level. 
There are many phenomenological and/or 
nonperturbative 
schemes aiming to explain the formation and value of QCD condensates 
(hence of QCD vacuum energy) that only partially can give insight into the 
problem \cite{QCD}. So a definite conclusion 
about this point needs 
further additional study. However this issue is not as urgent as the issue 
of zero point energies because the problem of zero point energies arises 
as soon as the fields are introduced (and quantized) even in the case of 
free fields while QCD vacuum is present only inside the hadrons and is not 
perfectly well understood. Another important issue to be studied in future 
is: Although I have shown that quantum fields do not induce non-vanishing 
vacuum energy at fundamental Lagrangian level (i.e. quantum zero modes do 
not contribute to vacuum energy) in the presence of metric reversal 
symmetry there is no guarantee of non-zero contributions to vacuum energy 
due higher dimension operators (than those of the fundamental Lagrangian). 
If this is the case the resulting vacuum energy due to quantum fields will 
be scale dependent through renormalization group equations. The most 
reasonable consequence of this, in turn, would be a time varying 
cosmological constant \cite{varying-CC}. Time varying cosmological 
constant scenarios together with  quintessence models have an additional 
virtue 
of explaining cosmic coincidence i.e. the energy density of matter and 
dark energy being in the same order of magnitude, that is not addressed 
by the scheme in this paper. All these studies must 
studied in future for a clearer picture of the cause and dynamics of the 
accelerated expansion of the universe.

\section{inducing a small cosmological constant by breaking the symmetry 
by a small amount}

We have seen that contribution of quantum fields to the 
energy-momentum tensor is always zero in the manifestation of 
signature reversal symmetry. However this is not true for 
classical fields. For example consider a classical field that depends 
only on extra dimensions and has a
Fourier expansion as in (\ref{cb6},\ref{cb9}). This field gives non-zero 
contribution to 4-dimensional cosmological constant (CC) after integration 
over 
extra dimensions. For example one may take
\begin{equation}
{\cal L}_{cl}\,=\,\alpha\,
v_{1,0}v_{0,1}
\cos{k\,y}\,\cos{k^\prime\,z} 
\label{e1}
\end{equation}
where $\alpha\,<<\,1$ is a constant that reflects that ${\cal L}_{cl}$ is 
small since it corresponds to the breaking of the 
$x^A\rightarrow\,i\,x^A$, 
$x^{A^\prime}\rightarrow\,i\,x^{A^\prime}$ 
symmetries separately by a small amount, and $v_{1,0}$, $v_{0,1}$ are some 
constants. If one takes the same space as in the section 4 and take 
$N=6$, $N^\prime=2$ (as before) then ${\cal L}_{cl}$ in (\ref{e1}) after 
integration over extra dimensions results in a 4-dimensional 
CC given by
\begin{equation}
\Lambda^{(4)}\,=\,\frac{3\alpha\,v_{1,0}v_{0,1}}{16}(LL^\prime)^2 
~~.\label{e2}
\end{equation}
For $\alpha\,v_{1,0}v_{0,1}\simeq\,1$ (\ref{e2}) results in the observed 
value of 
$\Lambda\simeq\,(10^{-3}eV)^4$ for $L$, $L^\prime$ in the millimeter scale 
and for $\alpha\,v_{1,0}v_{0,1}\simeq \,\frac{M_{ew}}{M_{pl}}\simeq 
10^{-17}$, 
for example,
$L(L^\prime)\,<\,10^{-7}m$. In any case a non-zero CC 
if exists is a classical phenomena in this scheme. Another point is that 
the energy density due to CC obtained in a way similar 
to (\ref{e2}) may be argued to be in the order of matter (ie. the usual 
matter plus dark matter) density since both are induced by matter 
Lagrangian that corresponds to breaking of the  
$x^A\rightarrow\,i\,x^A$, $x^{A^\prime}\rightarrow\,i\,x^{A^\prime}$ 
symmetries. However there is a difference between the two cases. The 
induction of $S_M$ corresponds to breaking the symmetry that corresponds 
to the simultaneous application of 
$x^A\rightarrow ix^A$ and 
$x^{A^\prime}\rightarrow ix^{A^\prime}$ while ${\cal L}_{cl}$ in 
Eq.(\ref{e1}) corresponds to breaking of $x^A\rightarrow ix^A$ and 
$x^{A^\prime}\rightarrow ix^{A^\prime}$ separately.

\section{Conclusion}

We have considered a space that is a sum of two $2(2n+1)$ dimensional 
spaces with $R^2$ gravity and metric reversal symmetry. The usual 
4-dimensional space is embedded in one of these subspaces. We have shown 
that the curvature sector reduces to the usual Einstein-Hilbert action, 
and the 4-dimensional energy-momentum tensor of matter fields generically 
mixes different Kaluza-Klein modes so that each homogeneous 
term contains at least one pair of off-diagonally coupled Kaluza-Klein 
modes. This, in turn, results in vanishing of the vacuum expectation value 
of the energy-momentum tensor of quantum fields. I have also shown that 
such a model is equivalent to a variation of Linde's model (where the 
universe consists of the usual universe plus a ghost one). There may be 
some relation between this scheme and the Pauli-Villars regularization 
scheme \cite{Pauli-Villars} ( that employs ghost-like auxiliary fields for 
regularization), and also between this scheme and Lee-Wick quantum theory 
\cite{Lee}. In my opinion all these points need further and detailed 
studies in future.

\begin{acknowledgments}
This work was supported in part by Scientific and Technical Research 
Council of Turkey under grant no. 107T235.
\end{acknowledgments}

\appendix
\section{Calculation of $S_{\phi k}$}
\begin{eqnarray}
S_{\phi k} &=& 
\int 
\,dV\,{\cal L}_{\phi k} 
\nonumber \\
&=&
\frac{1}{2}\int 
\,\sqrt{(-1)^S g}\sqrt{(-1)^{S^\prime} g^\prime} \,d^Dx\,d^Dx^\prime
[\frac{1}{2}g^{AB}\partial_A\phi\partial_B\phi\,+\,
\frac{1}{2}g^{A^\prime B^\prime}
\partial_{A^\prime}\phi\partial_{B^\prime}\phi] \nonumber \\
&=&
\frac{1}{2}\int\,d^4x\,dy_1dy_2dz_1dz_2\,\Omega_z^3\Omega_y\,
\{\Omega_z^{-1}[\eta^{\mu\nu}\partial_\mu\phi\partial_\nu\phi\, 
-\,(\frac{\partial\phi}{\partial y_1})^2
\,-\,(\frac{\partial\phi}{\partial y_2})^2] \nonumber \\
&&
\,-\,\Omega_y[
(\frac{\partial\phi}{\partial z_1})^2
\,+\,
(\frac{\partial\phi}{\partial z_2})^2]\,
\} \nonumber \\
&=&
\frac{1}{2}LL^\prime\int\,d^4x\,\int_0^L\int_0^{L^\prime}dydz\,
\cos^3
{k^\prime z}
\cos{k y}
\{\cos^{-1}{k^\prime z}
[\eta^{\mu\nu}\partial_\mu\phi\partial_\nu\phi\, 
\,-\,(\frac{\partial\phi}{\partial y})^2]
\,-\,\cos^{-1}{k y}
(\frac{\partial\phi}{\partial z})^2
\} \nonumber \\
\label{cba3}
\end{eqnarray}
First evaluate (\ref{cba3}) for (\ref{cb6})
\begin{eqnarray}
S_{Mk} &=& 
\frac{1}{2}LL^\prime\int\,d^4x\,\{
\eta^{\mu\nu}\sum_{n,m,r,s}\partial_\mu( 
\,\phi_{n,m}(x)\,)\,\partial_\nu(\,\phi_{r,s}(x)\,) \nonumber \\
&&\times\,\int_0^L\,dy\,
\cos{k y}\, 
\sin{(n\,k|y|)}
\sin{(r\,k|y|)}
\int_0^{L^\prime}\,dz\,
\cos^2{k^\prime z}
\sin{(m\,k^\prime|z|)})
\,\sin{(s\,k^\prime|z|)}) \nonumber \\
&&\,-k^2\,\,
\sum_{n,m,r,s}\,nr
\,\phi_{n,m}(x)\,\phi_{r,s}(x) 
\,\int_0^L\,dy\,
\cos{k y}\, 
\,\cos{(n\,k|y|)}
\cos{(r\,k|y|)} \nonumber \\
&&\times\,\int_0^{L^\prime}\,dz\,
\cos^2{k^\prime z}
\sin{(m\,k^\prime|z|)})
\,\sin{(s\,k^\prime|z|)}) \} \nonumber \\
&&\,-k^{\prime 2}\,\,
\sum_{n,m,r,s}\,ms
\,\phi_{n,m}(x)\,\phi_{r,s}(x) 
\,\int_0^L\,dy\,
\sin{(n\,k|y|)}
\sin{(r\,k|y|)} \nonumber \\
&&\times\,\int_0^{L^\prime}\,dz\,
\cos^3{k^\prime z}
\cos{(m\,k^\prime|z|)})
\,\cos{(s\,k^\prime|z|)}) \nonumber \\
&=&
\frac{1}{32}(LL^\prime)^2\int\,d^4x\,\{
\eta^{\mu\nu}\sum_{n,m,r,s}\partial_\mu( 
\,\phi_{n,m}(x)\,)\,\partial_\nu(\,\phi_{r,s}(x)\,) \nonumber \\
&&\times\,
(\delta_{n,r-1}+\delta_{n,r+1}-\delta_{n,-r-1}-
\delta_{n,1-r})(\delta_{m,s-2}+\delta_{m,s+2}-\delta_{m,-s-2}-
\delta_{m,2-s}+2\delta_{m,s}-2\delta_{m,-s}) \nonumber \\
&&\,-k^2\,\,
\sum_{n,m,r,s}\,nr
\,\phi_{n,m}(x)\,\phi_{r,s}(x) 
(\delta_{n,r-1}+\delta_{n,r+1}+\delta_{n,-r-1}+\delta_{n,1-r}) \nonumber \\
&&\times\,(\delta_{m,s-2}+\delta_{m,s+2}-\delta_{m,-s-2}-
\delta_{m,2-s}+2\delta_{m,s}-2\delta_{m,-s}) \nonumber \\
&&\,-\frac{1}{2}k^{\prime 2}\,\,
\sum_{n,m,r,s}\,ms
\,\phi_{n,m}(x)\,\phi_{r,s}(x) 
(
\delta_{n,r}-
\delta_{n,-r}) \nonumber \\
&&\times
\,(\delta_{m,s-3}+\delta_{m,s+3}+\delta_{m,-s-3}+\delta_{m,3-s}
+3\delta_{m,s-1}+3\delta_{m,s+1}+3\delta_{m,-s-1}+3\delta_{m,1-s})\}
\label{cba4} \\
&=&
\frac{1}{32}(LL^\prime)^2\int\,d^4x\,\{
\eta^{\mu\nu}\sum_{r,s}
\nonumber \\
&&
\partial_\mu 
[\,\phi_{r-1,s-2}(x)\,+\,\phi_{r-1,s+2}(x)
\,-\,\phi_{r-1,-s-2}(x)\,-\,\phi_{r-1,2-s}(x)
\,+\,2\phi_{r-1,s}(x)
\nonumber \\
&&-\,2\,\phi_{r-1,-s}(x) \,+\,\phi_{r+1,s-2}(x)\,+\,(\,\phi_{r+1,s+2}(x)
\,-\,\phi_{r+1,-s-2}(x)\,-\,\phi_{r+1,2-s}(x)\,+\,2\phi_{r+1,s}(x) 
\nonumber \\
&&-\,2\phi_{r+1,-s}(x)\,-\,\phi_{-r-1,s-2}(x)\,)\,-\,\phi_{-r-1,s+2}(x)\,+\,
\phi_{-r-1,-s-2}(x)\,+\,\phi_{-r-1,2-s}(x)
\nonumber \\
&&-\,2\phi_{-r-1,s}(x)\,+\,2\phi_{-r-1,-s}(x)
\,-\,\phi_{1-r,s-2}(x)\,-\,\phi_{1-r,s+2}(x)
\,+\,\phi_{1-r,-s-2}(x) \nonumber \\
&&+\,\phi_{1-r,2-s}(x)\,-\,2\phi_{1-r,s}(x)\,+\,2\phi_{1-r,-s}(x)
\,] \partial_\nu(\,\phi_{r,s}(x)\,) \nonumber \\
&&\,-k^2\,\,
\sum_{r,s}\,r
[\,(r-1)(\,\phi_{r-1,s-2}(x)\,-\,
\phi_{1-r,s-2}(x)\,)\,+\,(r-1)(\phi_{r-1,s+2}(x)
\,-\,\phi_{1-r,s+2}(x)\,)
\nonumber \\
&&-\,(r-1)(\,\phi_{r-1,-s-2}(x)\,-\,\phi_{1-r,-s-2}(x)\,)
\,-\,(r-1)(\,\phi_{r-1,2-s}(x)\,-\,\phi_{1-r,2-s}(x)\,) \nonumber \\
&&+\,2(r-1)(\,\phi_{r-1,s}(x)\,-\,\phi_{1-r,s}(x)\,)
\,-\,2(r-1)(\,\phi_{r-1,-s}(x)\,-
\,\phi_{1-r,-s}(x)\,) \nonumber \\
&&+\,(r+1)(\,\phi_{r+1,s-2}(x)\,-\,\phi_{-r-1,s-2}(x)\,)
\,+\,(r+1)(\,\phi_{r+1,s+2}(x)
\,-\,\phi_{-r-1,s+2}(x)\,) \nonumber \\
&&-\,(r+1)(\,\phi_{r+1,-s-2}(x)\,-\,\phi_{-r-1,-s-2}(x)\,)
\,-\,(r+1)(\,\phi_{r+1,2-s}(x)\,-\,\phi_{-r-1,2-s}(x)\,)\nonumber \\
&&+\,2(r+1)(\,\phi_{r+1,s}(x)\,-\,\phi_{-r-1,s}(x)\,)
\,-\,2(r+1)(\,\phi_{r+1,-s}(x)\,-\,\phi_{-r-1,-s}(x)\,)\,]
\phi_{r,s}(x) \nonumber \\
&&\,-\frac{1}{2}k^{\prime 2}\,\,\sum_{r,s}\,s
\,[\,(s-3)(\,\phi_{r,s-3}(x)\,-\,\phi_{r,3-s}(x)\,)
\,+\,(s+3)(\,\phi_{r,s+3}(x)\,-\,\phi_{r,-s-3}(x)\,) \nonumber\\
&&+\,3(s-1)(\,\phi_{r,s-1}(x)\,-\,\phi_{r,1-s}(x)\,)
\,+\,3(s+1)(\,\phi_{r,s+1}(x)\,-\,\phi_{r,-s-1}(x)\,)
\nonumber \\
&&+\,(3-s)(\,\phi_{-r,s-3}(x)\,-\,\phi_{-r,3-s}(x)\,)
\,+\,(s+3)(\,\phi_{-r,-s-3}(x)\,-\,\phi_{-r,s+3}(x)\,)
\nonumber \\
&&+\,3(1-s)(\,\phi_{-r,s-1}(x)\,-\,\phi_{-r,1-s}(x)\,)
\,-\,3(s+1)(\,\phi_{-r,s+1}(x)\,-\,\phi_{-r,-s-}(x)\,)\,]
\,\phi_{r,s}(x) \}
\label{cba5} 
\end{eqnarray}
where I have used (\ref{ba10}) and taken $y=y_2$, $z=z_2$, and 
$\phi_{mn}=\phi^{AA}_{mn}$ in (\ref{cb6}). 
After using the antisymmetry of 
$\phi^{AA}_{mn}$ 
under both of $n\rightarrow -n$
$m\rightarrow -m$, (\ref{cba5}) may be written in a simplified form as
(\ref{cb6}). In fact this result is essentially the same as those of 
$\phi_{AS}$,
$\phi_{SA}$,
$\phi_{SS}$ as explained after Eq.(\ref{cba6})

\section{Calculation of $S_{fk}$}
\begin{eqnarray}
S_{fk}
&=&
\frac{1}{2}LL^\prime\int\,d^4x\,\int_0^L\int_0^{L^\prime}dydz\,
\cos^3{k^\prime z}\cos{k y} \nonumber \\
&&\times\,[\,i
\bar{\psi}
\,(\cos{\frac{k\,z}{2}}\tau_3
\,+\,i\sin{\frac{k\,z}{2}}\tau_1)^{-1} 
\otimes\gamma^A
\partial_A\psi 
\,+\,i\bar{\psi}\,(\cos{\frac{k\,y}{2}}\tau_3
\,+\,i\sin{\frac{k\,y}{2}}\tau_1)^{-1} \otimes\gamma^{A^\prime}
\partial_{A^\prime}\psi \,] \nonumber \\
&=&
\frac{1}{2}LL^\prime\int\,d^4x\,\int_0^L\int_0^{L^\prime}dydz\,
[\,i\bar{\psi}
\,\cos^2{k^\prime z}\cos{k y}
(\cos{\frac{k\,z}{2}}\tau_3\,+\,i\sin{\frac{k\,z}{2}}\tau_1) 
\otimes\gamma^A\partial_A\psi \nonumber \\
&&+\,i\bar{\psi}\,
\,\cos^3{k^\prime z}
(\cos{\frac{k\,y}{2}}\tau_3\,+\,i\sin{\frac{k\,y}{2}}\tau_1) 
\otimes\gamma^{A^\prime}
\partial_{A^\prime}\psi \,] \nonumber \\
&=& 
\frac{1}{2}LL^\prime\int\,d^4x\,\{
\sum_{n,m,r,s} 
\,[\,
i\psi_{n,m}(x)\,\tau_3\otimes\gamma^\mu\partial_\mu(\,\psi_{r,s}(x)
\,\int_0^L\,dy\,
\cos{(\,k|y|)}
\exp{(\frac{i}{2}(r-n)\,k|y|)} \nonumber \\
&&\times\,\int_0^{L^\prime}\,dz\,
\cos^2{\,k^\prime|z|}
\cos{(\frac{1}{2}\,k^\prime|z|)}
\exp{(\frac{i}{2}(s-m)\,k^\prime|z|)}
\nonumber \\
&&-\,
\psi_{n,m}(x)\,\tau_1\otimes\gamma^\mu\partial_\mu(\,\psi_{r,s}(x)
\,\int_0^L\,dy\,
\cos{(\,k|y|)}
\exp{(\frac{i}{2}(r-n)\,k|y|)} \nonumber \\
&&\times\,\int_0^{L^\prime}\,dz\,
\cos^2{\,k^\prime|z|}
\sin{(\frac{1}{2}\,k^\prime|z|)}
\exp{(\frac{i}{2}(s-m)\,k^\prime|z|)}
\nonumber \\
&&-\,
\frac{1}{2}
\psi_{n,m}(x)\,(r-n)\,\tau_3\otimes\gamma^y \psi_{r,s}(x)
\,\int_0^L\,dy\,
\cos{(\,k|y|)}
\exp{(\frac{i}{2}(r-n)\,k|y|)} \nonumber \\
&&\times\,\int_0^{L^\prime}\,dz\,
\cos^2{\,k^\prime|z|}
\cos{(\frac{1}{2}\,k^\prime|z|)}
\exp{(\frac{i}{2}(s-m)\,k^\prime|z|)}
\nonumber \\
&&-i\frac{1}{2}
\,(r-n)\,
\psi_{n,m}(x)\,\tau_1\otimes\gamma^y
\,\psi_{r,s}(x)
\,\int_0^L\,dy\,\cos{(\,k|y|)}
\exp{(\frac{i}{2}(r-n)\,k|y|)} \nonumber \\
&&\times\,\int_0^{L^\prime}\,dz\,
\cos^2{\,k^\prime|z|}
\sin{(\frac{1}{2}\,k^\prime|z|)}
\exp{(\frac{i}{2}(s-m)\,k^\prime|z|)}
\nonumber \\
&&-\,
\frac{1}{2}
\frac{1}{2}\psi_{n,m}(x)\,(s-m)\,\tau_3\otimes\gamma^y \psi_{r,s}(x)
\,\int_0^L\,dy\,
\cos{(\frac{1}{2}\,k|y|)}
\exp{(\frac{i}{2}(r-n)\,k|y|)} \nonumber \\
&&\times\,\int_0^{L^\prime}\,dz\,
\cos^3{\,k^\prime|z|}
\exp{(\frac{i}{2}(s-m)\,k^\prime|z|)}
\nonumber \\
&&-i\,\frac{1}{2}
(s-m)\,
\psi_{n,m}(x)\,\tau_1\otimes\gamma^y
\,\psi_{r,s}(x)
\,\int_0^L\,dy\,\sin{(\frac{1}{2}\,k|y|)}
\exp{(\frac{i}{2}(r-n)\,k|y|)} \nonumber \\
&&\times\,\int_0^{L^\prime}\,dz\,
\cos^3{\,k^\prime|z|}
\exp{(\frac{i}{2}(s-m)\,k^\prime|z|)} \}
\nonumber \\
&=& 
\frac{1}{2}LL^\prime\int\,d^4x\,\{
\sum_{n,m,r,s} 
\,[\,
i\psi_{n,m}(x)\,\tau_3\otimes\gamma^\mu\partial_\mu(\,\psi_{r,s}(x)
\,\int_0^L\,dy\,
\cos{(\,k|y|)}
\cos{(\frac{1}{2}(r-n)\,k|y|)} \nonumber \\
&&\times\,\int_0^{L^\prime}\,dz\,
\cos^2{\,k^\prime|z|}
\cos{(\frac{1}{2}\,k^\prime|z|)}
\cos{(\frac{1}{2}(s-m)\,k^\prime|z|)}
\nonumber \\
&&-\,
\psi_{n,m}(x)\,\tau_1\otimes\gamma^\mu\partial_\mu(\,\psi_{r,s}(x)
\,\int_0^L\,dy\,
\cos{(\,k|y|)}
\cos{(\frac{1}{2}(r-n)\,k|y|)} \nonumber \\
&&\times\,\int_0^{L^\prime}\,dz\,
\cos^2{\,k^\prime|z|}
\sin{(\frac{1}{2}\,k^\prime|z|)}
\sin{(\frac{1}{2}(s-m)\,k^\prime|z|)}
\nonumber \\
&&-\,
\frac{1}{2}\psi_{n,m}(x)\,(r-n)\,\tau_3\otimes\gamma^y \psi_{r,s}(x)
\,\int_0^L\,dy\,
\cos{(\,k|y|)}
\cos{(\frac{1}{2}(r-n)\,k|y|)} \nonumber \\
&&\times\,\int_0^{L^\prime}\,dz\,
\cos^2{\,k^\prime|z|}
\cos{(\frac{1}{2}\,k^\prime|z|)}
\cos{(\frac{1}{2}(s-m)\,k^\prime|z|)}
\nonumber \\
&&+\,\frac{1}{2}
(r-n)\,
\psi_{n,m}(x)\,\tau_1\otimes\gamma^y
\,\psi_{r,s}(x)
\,\int_0^L\,dy\,
\cos{(\,k|y|)}
\cos{(\frac{1}{2}(r-n)\,k|y|)} \nonumber \\
&&\times\,\int_0^{L^\prime}\,dz\,
\cos^2{\,k^\prime|z|}
\sin{(\frac{1}{2}\,k^\prime|z|)}
\sin{(\frac{1}{2}(s-m)\,k^\prime|z|)}
\nonumber \\
&&-\,
\frac{1}{2}\psi_{n,m}(x)\,(s-m)\,\tau_3\otimes\gamma^y \psi_{r,s}(x)
\,\int_0^L\,dy\,
\cos{(\frac{1}{2}\,k|y|)}
\cos{(\frac{1}{2}(r-n)\,k|y|)} \nonumber \\
&&\times\,\int_0^{L^\prime}\,dz\,
\cos^3{\,k^\prime|z|}
\cos{(\frac{1}{2}(s-m)\,k^\prime|z|)}
\nonumber \\
&&+\,(s-m)\frac{1}{2}
\,
\psi_{n,m}(x)\,\tau_1\otimes\gamma^y
\,\psi_{r,s}(x)
\,\int_0^L\,dy\,
\sin{(\frac{1}{2}\,k|y|)}
\sin{(\frac{1}{2}(r-n)\,k|y|)} \nonumber \\
&&\times\,\int_0^{L^\prime}\,dz\,
\cos^3{\,k^\prime|z|}
\cos{(\frac{1}{2}(s-m)\,k^\prime|z|)} \}~~.
\end{eqnarray}
After integration over y and z this equation results in (\ref{cbc7}).

\section{Expilicit forms of $C_K$, {\small K =1,2,3,.....,12}}
After inserting Eqs. (\ref{lba10}) and (\ref{lba11}) (in the light of Eqs. 
(\ref{lba7}) and (\ref{lba8}) ) into Eq.(\ref{cba3}) and integrating over 
the extra dimensions it should be equal to (\ref{cba6}). Hence after 
comparing the result of the integration with Eq.(\ref{cba6}) we obtain the 
following results for the constants, 
\begin{eqnarray} 
2C_1C_2&=& \sum_{j,l}\{
\frac{|2j-1|\,|2l-1|}{((2j+1)^2+1)((2l+1)^2+1)}\,
[\,\frac{|2j-2|\,|2l-3|}{((2j)^2+1)((2l-1)^2+1)}\nonumber \\
&&+\,
\frac{|2j-2|\,|2l+1|}{((2j)^2+1)((2l+3)^2+1)}\,+\,
2\frac{|2j-2|\,|2l-1|}{(2j)^2(2l+1)^2}\nonumber \\
&&+\,\frac{|2j|\,|2l-1|}{((2j+2)^2+1)((2l-1)^2+1)}\,+\,
\frac{|2j|\,|2l+1|}{(2j+2)^2(2l+3)^2}\nonumber \\
&&+\,
2\frac{|2j|\,|2l-1|}{((2j+2)^2+1)((2l+1)^2+1)}\,]\,\} \nonumber \\
2C_3C_4&=&
\sum_{j,l}\{\,
\frac{|2j-1|\,|2l-2|}{((2j+1)^2+1)((2l)^2+1)}\,
[\,\frac{|2j-2|\,|2l-4|}{((2j)^2+1)((2l-2)^2+1)}\nonumber \\
&&+\,
\frac{|2j-2|\,|2l|}{((2j)^2+1)((2l+2)^2+1)}\,+\,
2\frac{|2j-2|\,|2l-2|}{((2j)^2+1)((2l)^2+1)}\nonumber \\
&&+\,\frac{|2j|\,|2l-4|}{((2j+2)^2+1)((2l-2)^2+1)}\,+\,
\frac{|2j|\,|2l|}{((2j+2)^2+1)((2l+2)^2+1)}\nonumber \\
&&+\,
2\frac{|2j|\,|2l-2|}{((2j+2)^2+1((2l)^2+1)}\,]\,\} \nonumber \\
2C_5C_6&=&
\sum_{j,l}\{\,(2j)
\frac{|2j-2|\,|2l-1|}{((2j)^2+1)(2l+1)^2+1)}\,
[\,
(2j-1)(\,\frac{|2j-3|\,|2l-3|}{((2j-1)^2+1)((2l-1)^2+1)}\nonumber \\
&&+\,
\frac{|2j-3|\,|2l+1|}{((2j-1)^2+1)((2l+3)^2+1)}\,+\,
2\frac{|2j-3|\,|2l-1|}{((2j-1)^2+1)((2l+1)^2+1)}\,)\nonumber \\
&&+\,(2j+1)(\,
\frac{|2j-1|\,|2l-3|}{((2j+1)^2+1)((2l-1)^2+1)}\nonumber \\
&&+\,
\frac{|2j-1|\,|2l+1|}{((2j+1)^2+1)((2l+3)^2+1)}\,+\,
2\frac{|2j-1|\,|2l-1|}{((2j+1)^2+1)((2l+1)^2+1)}\,)\,]\,\} \nonumber \\
2C_7C_8&=&
\sum_{j,l}\{\,
(2j+1)(\,\frac{|2j-1|\,|2l-2|}{((2j+1)^2+1)((2l)^2+1)}\,
[\,
(2j)(\,\frac{|2j-2|\,|2l-4|}{((2j)^2+1)((2l-2)^2+1)}\nonumber \\
&&+\,
\frac{|2j-2|\,|2l|}{((2j)^2+1)((2l+2)^2+1)}\,+\,
2\frac{|2j-2|\,|2l-2|}{((2j)^2+1)((2l)^2+1)}\nonumber \\
&&+\,(2j+2)(\,
\frac{|2j|\,|2l-4|}{((2j+2)^2+1)((2l-2)^2+1)}\,+\,
\frac{|2j|\,|2l|}{((2j+2)^2+1)((2l+2)^2+1)}\nonumber \\
&&+\,
2\frac{|2j|\,|2l-2|}{((2j+2)^2+1)((2l)^2+1)}\,]\,\} \nonumber \\
2C_9C_{10}&=&
\sum_{j,l}\{\,
\,(2l)(\,\frac{|2j-1|\,|2l-2|}{((2j+1)^2+1)((2l)^2+1)}\,
[\,(2l-3)
\frac{|2j-1|\,|2l-5|}{((2j+1)^2+1)((2l-3)^2+1)}\nonumber \\
&&+\,
(2l+3)\frac{|2j-1|\,|2l+1|}{((2j+1)^2+1)((2l+3)^2+1)}\nonumber \\
&&+\,
3(2l-1)\frac{|2j-1|\,|2l-3|}{((2j+1)^2+1)((2l-1)^2+1)}\,+\,
3(2l+1)\frac{|2j-1|\,|2l-1|}{((2j+1)^2+1)((2l+1)^2+1)}\,]\,\} \nonumber \\
2C_{11}C_{12}&=&
\sum_{j,l}\{\,
(2l+1)\frac{|2j-1|\,|2l-2|}{((2j)^2+1)((2l+1)^2+1)}\,
[\,
(2l-2)\frac{|2j-2|\,|2l-4|}{((2j)^2+1)((2l-2)^2+1)}\nonumber \\
&&+\,
(2l+4)\frac{|2j-2|\,|2l+2|}{((2j)^2+1)((2l+4)^2+1)}\,+\,
3(2l)\frac{|2j-2|\,|2l-2|}{((2j)^2+1)((2l)^2+1)}\nonumber \\
&&+\,
3(2l+2)\frac{|2j-2|\,|2l|}{((2j)^2+1)((2l+2)^2+1)}\,]\,\}~~.
\label{cba13}
\end{eqnarray}

\bibliographystyle{plain}

\end{document}